\documentclass{aa}  
\usepackage{graphicx}
\usepackage{txfonts}
\usepackage{comment}
\usepackage{color}
\usepackage{xcolor}
\usepackage{url}
\usepackage{hyperref}
\usepackage{multirow}
\usepackage{float}
\usepackage{orcidlink}
\usepackage{soul}
\usepackage{lipsum} 

\definecolor{myblue}{HTML}{6a359c}
\definecolor{myviolet}{HTML}{756BB1}
\definecolor{mygreen}{HTML}{7FCDBB}
\definecolor{myyellow}{HTML}{FFFDD1}

\hypersetup{
	colorlinks,
	linkcolor={myblue},
	citecolor={myblue},
	urlcolor={myblue}}

\usepackage{comment} %
\usepackage{color} %
\usepackage{amsmath} %
\usepackage{ulem} %
\usepackage{url}
\usepackage{orcidlink}
\newcommand{\Mjup}{\mbox{M$_{\mathrm{jup}}$}}
\newcommand{\Msun}{\mbox{M$_{\odot}$}}
\newcommand{\Mstar}{\mbox{M$_{\star}$}}
\newcommand{\Rstar}{\mbox{R$_{\star}$}}

\newcommand{\Teff}{\mbox{T$^{\text{eff}}_\star$}}

\newcommand{\Rsun}{\mbox{R$_{\odot}$}}

\newcommand{\SigmaG}{\Sigma_{\rm g}}

\defcitealias{Ronco2024}{PI}

\begin{document} 

\title {Planet formation around intermediate-mass stars.
}
\subtitle{II. From dust growth and pebble accretion to giant planet populations}

\titlerunning{Planet formation around intermediate-mass stars}
\authorrunning{Ronco et al.}

\author{
M. P. Ronco\orcidlink{0000-0003-1385-0373}\inst{1}, 
M. R. Schreiber\orcidlink{0000-0003-3903-8009}\inst{2},
O. M. Guilera\orcidlink{0000-0001-8577-9532}\inst{1},
M. Jones\orcidlink{0009-0000-9059-2430
}\inst{3},
E. Villaver\orcidlink{0000-0003-4936-418}\inst{4}\inst{5} 
}
\offprints{mpronco@fcaglp.unlp.edu.ar}
\institute{
Instituto de Astrof\'{\i}sica de La Plata, CCT La Plata-CONICET-UNLP, Paseo del Bosque S/N (1900), La Plata, Argentina,
\and
Departamento de F{\'i}sica, Universidad T\'ecnica Federico Santa Mar\'ia, Av. España 1680, Valpara{\'i}so, Chile, 
\and
European Southern Observatory, Alonso de Córdova 3107, Vitacura, Casilla 19001, Santiago, Chile,
\and
Instituto de Astrofísica de Canarias, 38200 La Laguna, Tenerife, Spain,
\and 
Universidad de La Laguna (ULL), Astrophysics Department, 38206 La Laguna, Tenerife, Spain
}

\date{}

\abstract{While theoretical models of planet formation have primarily focused on solar and low-mass stars, the intermediate-mass regime remains less explored from a modeling perspective. On the observational side, radial-velocity (RV) surveys show that giant planet occurrence peaks around $1.7 \mathrm{M}_\odot$ before declining. In this higher-mass regime, this dependence may reflect changes in both, the protoplanetary disk lifetimes and evolutionary pathways, caused by the evolving X-ray and far-ultraviolet (FUV) emission of pre-main-sequence (PMS) stars, which regulate disk dispersal through photoevaporation.}
{We investigate how protoplanetary disk evolution, driven by pre-main-sequence (PMS) stellar evolution, shapes the properties of giant planets emerging around 1–4 $\mathrm{M}_\odot$ stars. Through a systematic parameter study, our goal is to characterize the theoretical gas giant population that naturally forms in this regime, and subsequently evaluate how this population compares with current observational trends.
}
{Using the 1D+1D PLANETALP code, we simultaneously model gas disk evolution, dust growth, pebble and gas accretion, and planet migration, including thermal torques. Disks evolve through viscous accretion and time-dependent X-ray and FUV photoevaporation. We analyze a grid of 5400 planet formation simulations spanning stellar mass from $1$–$4$~M$_{\odot}$ and varying key disk parameters such as metallicity, initial disk mass and viscosity, and the onset of planet formation.}
{Our simulations show giant planet occurrence rising between $1$ and $3~\mathrm{M}_\odot$, peaking near $3~\mathrm{M}_\odot$, above the $\sim 1.7~\mathrm{M}_\odot$ peak inferred from RV surveys. Although intense X-ray and FUV emission progressively reduces disk lifetimes as stellar mass increases, giant planet formation is only effectively suppressed beyond $3~\mathrm{M}_\odot$. However, this critical mass threshold can strongly depend on the adopted photoevaporation prescriptions. Occurrence also increases with stellar metallicity, consistent with observed trends. Most simulated giant planets reside beyond $5$ au, a consequence of the combined effects of thermal torques and the specific disk evolutionary pathways, limiting the overlap with the populations probed by RV surveys and motivating comparisons with direct-imaging surveys.}
{PMS-dependent photoevaporation plays a central role in shaping the stellar-mass dependence of giant planet formation. Our models qualitatively recover the rise in occurrence toward intermediate stellar masses and the subsequent steep decline. The predicted peak occurs at a higher mass than observed but more quantitative comparisons between model predictions and observations require population synthesis studies.}

\keywords{Stars: evolution; Protoplanetary disks; planets and satellites: formation; Planets and satellites: gaseous planets}

\maketitle
%
\section{Introduction}
\label{Intro}

Between 10 and 15\% of solar-type stars host Jupiter-like planets \citep[e.g.,][]{Cumming2008, Mayor2011}. The occurrence rate of these gas giants has been found to strongly correlate with the host star metallicity \citep[e.g.][]{fischer2005, Johnson2010, Maldonado2025}, a relationship whose chemical signature itself depends on the stellar mass \citep{Maldonado2026}. Radial velocity (RV) surveys have shown that the fraction of stars hosting such planets increases with stellar mass over the range of approximately $0.5{-}2\,\mathrm{M}_\odot$ \citep[e.g.,][]{Johnson2010, Reffert2015, Jones2016, Borgniet2019}. However, this trend does not continue indefinitely. Combining data from the EXPRESS \citep{Jones2011}, PPPS\citep{Wittenmyer2011}, and Lick Giant Star surveys \citep{Frink2001}, \citet{Wolthoff2022} placed the peak occurrence rate at approximately $1.7\,\mathrm{M}_\odot$ and showed that the frequency of giant planets likely decreases substantially in the $2-4\,\mathrm{M}_\odot$ range.

While RV surveys probe the inner planetary architectures, direct-imaging surveys suggest that the enhanced occurrence of massive planets around intermediate-mass stars extends to much wider orbital separations. Observations from surveys such as GPIES \citep{Nielsen_2019}, SHINE \citep{Vigan_2021, Chomez2025}, and BEAST \citep{Delorme2024} support the emerging picture that intermediate-mass stars (including BA and B-type stars) are particularly efficient hosts of massive planets and substellar companions at tens or hundreds of astronomical units. Understanding the formation efficiency in these systems is also crucial for explaining the high occurrence of metal-polluted white dwarfs (with main-sequence progenitors in the $1.5-3\,\mathrm{M}_\odot$ range), which require the existence of planetary systems to drive the observed pollution \citep{Koester2014}.

The physical origin of the dependence of planet formation on stellar-mass is most likely connected to the initial properties and subsequent evolution of protoplanetary disks around intermediate-mass pre-main-sequence stars, where planets are expected to form. The central stars are typically divided into luminous Herbig Ae/Be stars with radiative envelopes and their convective intermediate-mass T\,Tauri progenitors \citep{Brittain_2023}. Observations indicate that Herbig Ae/Be stars experience substantial accretion, with rates significantly higher than those of lower-mass classical T\,Tauri stars \citep{Fairlamb_2015, Grant_2022} and their T\,Tauri progenitors \citep{Brittain_2026}. Furthermore, a transition in the star–disk interaction appears to occur around $4\,\mathrm{M}_\odot$, where the relation between accretion rate and stellar mass changes \citep{Wichittanakom_2020}. Simultaneously, these actively evolving disks provide direct evidence of ongoing planet formation. High-resolution ALMA and infrared observations reveal rings, gaps, and spirals \citep[e.g.,][]{Andrews_2020, Brittain_2023}. Observations of Herbig stars reveal complex structures and embedded giant planet candidates in systems like HD 100546 and AB Aur \citep{Quanz_2013, Currie_2022,Woelfer_2025}, alongside indirect kinematic evidence of forming planets in HD 163296 \citep{Pinte_2018}.

Forming giant planets requires a substantial gas reservoir. Shorter disk lifetimes, potentially driven by the evolving high-energy irradiation of the host star, leave less time for planetary cores to grow and undergo runaway gas accretion. \citet{KennedyKenyon2008} showed that disk lifetimes can influence the stellar-mass dependence of giant-planet occurrence. Through planet population synthesis based on planetesimal accretion, \citet{Alibert2011} subsequently showed that shorter disk lifetimes above $\sim 1.5\,$M$_\odot$ can explain the scarcity of short-period planets around $\sim 2\,$M$_\odot$ stars. 

More recently, \citet{johnston2024, johnston_2025, johnston_2026} investigated giant-planet formation around intermediate-mass stars using a pebble-accretion model, progressively refining their treatment of the disk environment. In their latest work, specifically aiming to reproduce the stellar-mass dependence of the giant planet occurrence rate \citep{Reffert2015,Jones2015,Wolthoff2022}, they performed a population synthesis study combining evolving PMS luminosities, stellar-mass-dependent disk accretion rates, and time-independent (stellar mass dependent) X-ray photoevaporation rates. Their models successfully reproduced the observed decline in giant-planet occurrence toward higher stellar masses, suggesting that the stellar-mass dependence of planet formation is strongly driven by the evolving disk environment. However, the high-energy irradiation experienced by a protoplanetary disk depends on more than just the stellar mass and luminosity; in particular, both X-ray and FUV luminosities change with stellar age, altering the strength and relative importance of the different photoevaporation mechanisms over the disk lifetime.

\citet{Kunitomo2021}, extending the models of \citet{Gorti2009}, showed that the dominant photoevaporation mechanism also depends strongly on stellar mass: while X-ray irradiation provides the main contribution for stars below $\sim2.5$\,\Msun, FUV irradiation becomes increasingly important for intermediate-mass stars, approximately $2.5
-5$\,\Msun. Importantly, the corresponding irradiation fields evolve with stellar age, implying that the relative contribution of X-ray and FUV photoevaporation can change during the lifetime of a disk. 

In \citet{Ronco2024} (hereafter \citetalias{Ronco2024}) we incorporated these time-dependent, mass-dependent photoevaporation processes into a detailed model of protoplanetary disk evolution and found that the transition from X-ray- to FUV-dominated photoevaporation is accompanied by qualitatively different disk-evolution pathways. In particular, the conventional inside-out clearing of the disk changes around $1.5-2$\,\Msun~while more massive stars can undergo even more complex evolutionary pathways.
These results raise the question of how planet formation proceeds when the time-dependent, mass-dependent evolution of the stellar irradiation field and the resulting complex disk-evolution pathways are taken into account. 

To address this, we couple the gas disk pathways from \citetalias{Ronco2024} with a comprehensive model of dust growth and evolution, pebble and gas accretion, and planet migration (including thermal torques). Rather than a standard population synthesis, we systematically explore a discrete parameter space for intermediate-mass stars (1–4 $\mathrm{M}_\odot$) to investigate the naturally emerging planet populations. By comparing our results with current observational trends, we aim to identify the initial conditions that could best reproduce giant planet occurrences and understand the physical mechanisms driving this stellar-mass dependence.

\section{Model Implementation}

The results presented in this work were obtained using PLANETALP, a 1D+1D code that models protoplanetary disk evolution and planetary growth \citep{Ronco2017,Guilera2019,Guilera2020,Venturini2020SE,Venturini2020Letter,Guilera2021,Venturini2024}. As described in \citetalias{Ronco2024}, the code fully solves the vertical structure equations for each radial bin \citep{Guilera2017,Guilera2019}. While \citetalias{Ronco2024} focused exclusively on the gas dynamics driven by early stellar evolution, in this work we activate the solid component and planetary modules to compute the simultaneous evolution of gas, dust, and planet formation. The aim of this second paper is to predict the main characteristics of the planet population formed within protoplanetary disks that have been affected by the early effects of stellar evolution around intermediate-mass stars. 

\subsection{Gas disk evolution}

To model the vertical structure and evolution of the gas disk, we adopt the framework developed in \citetalias{Ronco2024} (see their section 2 for details). The disk is assumed to be axisymmetric, geometrically thin, irradiated by the central star, and in hydrostatic equilibrium. For each radial bin, the vertical structure is obtained by solving the equations of hydrostatic equilibrium, radiative energy transport, and the temperature gradient, assuming a kinematic viscosity governed by the standard $\alpha$-prescription. The stellar radius and effective temperature used to determine this local vertical structure are kept fixed at their birthline values from \citet{Kunitomo2021}; that is, we neglect their temporal evolution. This simplification does not affect the resulting disk structure as shown in Appendix A and B of \citetalias{Ronco2024}.

In contrast, when simulating the long-term radial evolution of the gaseous disk, we explicitly account for the time evolution of the relevant stellar properties, including the total luminosity, effective temperature, radius, and high-energy emission. The evolution of the gas surface density is governed by the classical radial diffusion equation (eq.~3 in \citetalias{Ronco2024}), accounting for both viscous accretion and mass loss through photoevaporation. At each radius, the photoevaporative mass-loss rate is taken as the larger of the rates produced by X-ray and far-ultraviolet (FUV) irradiation. The X-ray photoevaporation rate is parameterized in terms of the evolving stellar X-ray luminosity. Similarly, the FUV photoevaporation rate scales with the total FUV luminosity, including contributions from the stellar photosphere, chromosphere, and accretion. The time-dependent stellar quantities entering these prescriptions are taken from table~1 of \citet{Kunitomo2021}. 

\subsection{Dust growth and evolution}

The dust growth and evolution, briefly summarized in this section, is computed as a simplified version of the one used in \citet{Guilera2020} and \citet{Venturini2020SE,Venturini2020Letter,Venturini2024} which is based on the results presented by \citet{Birnstiel11, Birnstiel12} and \citet{Drazkowska16}.
The maximum size of the dust particles at each radial bin, which is limited by dust coagulation, radial drift and fragmentation, is given by:
\begin{equation}
  r_{\text{d}}^{\text{max}}(t)= \min(r_{\text{d}}^0 \, \exp(t/\tau_{\text{growth}}), \, r_{\text{drift}}^{\text{max}}, \, r_{\text{frag}}^{\text{max}}),
  \label{eq1-sec2-1-0}
\end{equation}
where $r_{\text{d}}^0= 1~\mu\text{m}$ is the initial dust size.
At each radial bin and each time step there is a dust/pebble size distribution between 1 $~\mu\text{m}$ and $r_{\text{d}}^{\text{max}}$ given by Eq.(\ref{eq1-sec2-1-0}).
The collisional growth timescale ($\tau_{\text{growth}}$) is related to the dust-to-gas ratio $Z$ and the Keplerian frequency ($\Omega_{\text{k}}$): 
\begin{equation}
  \tau_{\text{growth}}= \dfrac{1}{Z\Omega_{\text{k}}}. 
  \label{eq2-sec2-1-0}
\end{equation}
The maximum size of dust particles limited by radial drift is given by 
\begin{equation}
  r_{\text{drift}}^{\text{max}}= f_{\text{d}} \dfrac{2\Sigma_{\text{d}}v_{\text{k}}^2}{\pi \rho_{\text{d}} c_s^2} \left| \dfrac{d\,\ln P_{\text{g}}}{d\,\ln r} \right|^{-1}, 
  \label{eq:drift}
\end{equation}
where $f_{\text{d}}= 0.55$ \citep{Birnstiel12}, $v_{\text{k}}$ is the Keplerian velocity, $P_{\text{g}}$ the gas pressure, and $\rho_{\text{d}}$ is the mean dust density, 
taking values of $\rho_{\text{d}}$ =$3~\text{g}/\text{cm}^3$ and $\rho_{\text{d}}$=$1~\text{g}/\text{cm}^3$ inside and outside the ice line, respectively.
These values assume pure silicate grains within the ice line and an ice-silicate mixture beyond it  \citep{Drazkowska17}. 
The maximum size of dust particles limited by fragmentation is: 
\begin{equation}
  r_{\text{frag}}^{\text{max}}= f_{\text{f}} \dfrac{2 \SigmaG v_{\text{th}}^2}{3\pi \rho_{\text{d}} \alpha_t c_s^2}, 
  \label{eq:frag}
\end{equation}
where $f_{\text{f}}= 0.37$ \citep{Birnstiel11}, and we use a fragmentation threshold velocity $v_{\text{th}}$= $1~\text{m}/\text{s}$ for silicate dust and $v_{\text{th}}$=$10~\text{m}/\text{s}$ for icy dust \citep{Drazkowska17}.

As the dust is strongly coupled to the gas we compute the time evolution of the dust surface density, $\Sigma_{\text{d}}$, by solving an advection-diffusion equation, 
\begin{equation}
  \frac{\partial}{\partial t} \left(\Sigma_{\text{d}}\right) + \frac{1}{r} \frac{\partial}{\partial r} \left( r \,\overline{v}_{\text{drift}} \,\Sigma_{\text{d}} \right) - \frac{1}{r} \frac{\partial}{\partial r} \left[ r \,D^{*} \,\SigmaG   \frac{\partial}{\partial r} \left( \frac{\Sigma_{\text{d}}}{\SigmaG} \right) \right] = \dot{\Sigma}_{\text{d}},  
  \label{eqAdvDif}
\end{equation}
where $D^{*}= \nu / ( 1 + \text{St}^2 )$ is the dust diffusivity \citep{YoudinLithwick2007}, and $\text{St}= \pi \rho_{\text{d}} {r}_{\text{d}} / 2 \SigmaG$ the mass weighted mean Stokes number of the dust size distribution. 
The mass weighted mean radius of the dust size distribution ($r_{\text{d}}$) is given by
\begin{equation}
  r_{\text{d}}= \dfrac{ \sum_i \epsilon_i r_{\text{d}}^i }{ \sum_i \epsilon_i },  
  \label{eq2-sec2-1}
\end{equation}
where $r_{\text{d}}^i$ represents the radius of the dust particles of species $i$, and $\epsilon_i= \rho^{i}_{\text{d}}/\rho_{\text{g}}$ the ratio between the volumetric dust density of species $i$ and the volumetric gas density \citep[see][]{Guilera2020}. 
The mean drift velocities of the pebbles population ($\overline{v}_{\text{drift}}$) are computed following \citet{Weiden77}. The term $\dot{\Sigma}_{\text{d}}$ represents the sink term due to pebble sublimation, assuming that pebbles drifting inward from the outer disk lose 50\% of their mass upon crossing the water ice line.

The initial dust surface density is given by $\Sigma_{\text{d}}(r) = \eta_{\text{ice}} Z \SigmaG(r)$ 
where $\SigmaG(r)$ is the initial gas surface density \citepalias[see][their eq.\,7]{Ronco2024}. 
The sublimation of water-ice is taken into account as 
$\eta_{\text{ice}}$=1/2 inside and $\eta_{\text{ice}}$=1 outside the ice-line \citep{Lodders09}. 
The dependence of the dust to gas mass ratio on metallicity of the host star ($[\text{Fe}/\text{H}]$) is assumed 
to be given by $Z = Z_0 10^{[\text{Fe}/\text{H}]}$, with $Z_0 = 0.0187$ the primordial abundance of heavy elements in the Sun \citep{Lodders2025}. 

\subsection{Planet formation model}
\label{sec:planet_formation_model}

We compute the growth of a single moon-mass protoplanetary embryo, initially located at different positions within the disk. The planet grows by pebble accretion following \citet{LambrechtsJohansen2014} and \citet{Brasser2017} \citep[see sect. 2.3 in][for details]{Venturini2020SE}. This phase continues until the embryo perturbs the disk, creating a pressure bump that halts pebble accretion. Following \citet{Lambrechts2014}, this pebble isolation mass scales with the mass of the star \citep[see also][]{LiuBeibei2019} and is given by:
\begin{equation}\label{MisoEq}
M_{\text{iso}} = 20 \, M_{\oplus}\, \left(\frac{H_{\text{g}}/r}{0.05} \right)^3 \frac{M_\star}{M_\odot}
\end{equation}
with $H_{\text{g}}$ representing the scale height of the disk.

After reaching the pebble isolation mass, we assume no further solid accretion and the planet grows solely via gas accretion \citep{Ikoma2000}. As the protoplanet evolves, its required thermal gas accretion rate can exceed what the disk is able to provide. When this happens, the planet detaches from the disk and its growth is strictly limited by the disk's viscous accretion. Finally, once the planet grows massive enough to open a gap in the disk \citep{Crida2006}, the gas accretion is significantly reduced, as suggested by \citet{Tanigawa07}.

Our planets are able to migrate along the disk as they grow. For low- and intermediate-mass planets (type I migration regime), the classical Lindblad ($\Gamma_L$) and corotation ($\Gamma_C$) torques are computed adopting the prescriptions derived by \citet{JM17}. We also take into account the contribution of thermal torques \citep{Masset2017, VelascoRomero2020}. These thermal torques ($\Gamma_{\text{thermal}}$), particularly the heating torques \citep{Benitez-llambay2015}, tend to be positive, allowing the planet to migrate outward.

Once a planet grows massive enough to carve a gap in the gas disk, it enters the Type II migration regime. We unify the Type I and Type II migration regimes following the prescription of \citet{Kanagawa2018}. Derived from 2D hydrodynamic simulations, their empirical formula smoothly bridges the torques exerted on low-mass, intermediate, and massive gap-opening planets. Thus, the total torque acting on the planet is computed as:
\begin{equation}
    \Gamma = \frac{\Gamma_L + \Gamma_C \exp(-K/K_{\mathrm{t}})}{1 + 0.04K} + \Gamma_{\rm thermal},
    \label{eq:torques}
\end{equation}
with
\begin{equation}
    K = \left( \frac{M_{\mathrm{p}}}{M_*} \right)^2 \left( \frac{H_{\mathrm{g,p}}}{a_{\mathrm{p}}} \right)^{-5} \alpha^{-1},
    \label{eq:k-parameter}
\end{equation}
where $K_{\mathrm{t}}= 20$, $M_{\mathrm{p}}$ and $a_{\mathrm{p}}$ are the planet mass and planet location, respectively, and $H_{\mathrm{g,p}}$ the disk scale height at such location. We note here that for low-mass planets $K \ll 20$ and $0.04K \ll 1$, recovering the classical type I migration recipes \citep{Tanaka02}. As planets grow, corotation and thermal torques vanish and planets migrate according to the results from \citet{Kanagawa2018}.   

\section{Initial conditions}
\label{IC}

\begin{table*}
\caption{Star and disk initial parameters for the added case of $4\Msun$. $\Rstar$ and $\Teff$ refer to the radius and effective temperature of the star at the birthline. $R_{\text{int}}$ and $R_{\text{c}}$ represent the inner border and the characteristic radius of the disk, respectively. $\alpha$ is the viscosity parameter and $\text{M}_{\text{d}}$ the mass of the disk. The last column refers to $\tau$, the disk dissipation timescale, which is not an initial condition but a result of our simulations.}
\begin{center}
  \begin{tabular}{|c|c|c|c|c|c|c|c|c|}
    \hline
       Sim & $\Mstar$ $[\Msun]$ & $\Rstar$ $[\Rsun]$ & $\Teff$ [K] & $R_{\text{int}}$ [au] & $R_{\text{c}}$ [au] & 
       $\alpha$ & $\text{M}_{\text{d}}$ [$\Msun$] & $\tau$ [Myr]\\
      \hline
       S46 & 4.0 & 4.571 & 5415 & 0.158 & 95.13 & $4x10^{-3}$ & 0.12 & 0.75 \\
       S47 & 4.0 & 4.571 & 5415 & 0.158 & 95.13 & $4x10^{-3}$ & 0.24 & 0.94 \\
       S48 & 4.0 & 4.571 & 5415 & 0.158 & 95.13 & $4x10^{-3}$ & 0.40 & 0.99 \\
       S49 & 4.0 & 4.571 & 5415 & 0.158 & 95.13 & $4x10^{-4}$ & 0.12 & 7.07 \\
       S50 & 4.0 & 4.571 & 5415 & 0.158 & 95.13 & $4x10^{-4}$ & 0.24 & 7.53 \\
       S51 & 4.0 & 4.571 & 5415 & 0.158 & 95.13 & $4x10^{-4}$ & 0.40 & 7.85 \\     
      \hline
   \end{tabular}
  \end{center}
  \label{Tabla-4Msun}
\end{table*}

To investigate the impact of time-dependent photoevaporation on protoplanetary disk evolution and planet formation around intermediate-mass stars, we adopt the initial disk and stellar parameters considered in \citetalias{Ronco2024} (see their Table 1) which are summarized here. We explore the same stellar masses, $1\Msun,~ 1.5\Msun,~2\Msun,~2.5\Msun$ and $3\Msun$, and now also include $4\Msun$, the disk and stellar parameters for this new case are given in Table \ref{Tabla-4Msun} ( a complement to Table 1 in \citetalias{Ronco2024}). The inner edge and the characteristic cut-off radius of the
disk were scaled with the mass of the central star as $R_\text{in} =0.1(M_\star/M_\odot)^{1/3}$ and $R_\text{c} =R^0_\text{c}
(M_\star/M_\odot)^{0.625}$ following \citet{Burn2021}, with $R^0_\text{c}
= 40$ au \citep{Andrews2010}. The initial
disk mass scales linearly with stellar mass, covering scenarios
for initially low-mass (3\%), intermediate (6\%), and massive
(10\%) disks. Finally, the turbulent viscosity parameter is also scaled with stellar mass as $\alpha = \alpha_0(M_\star/M_\odot)$, exploring high ($\alpha_0 = 10^{-3}$), and
low viscosity ($\alpha_0 = 10^{-4}$) environments. 

For each combination of disk mass and viscosity, we account for five values of stellar metallicity given by $[\text{Fe}/\text{H}]=-0.5$, -0.2, 0.0, 0.2 and 0.5, which lead to values of dust-to-gas ratios of $Z=0.006, 0.012, 0.0187, 0.03$ and $0.06$, respectively.
We compute the growth of 10 individual embryos per disk with an initial mass equal to that of the Moon ( 0.01~$M_\oplus$). The embryos are logarithmically spaced in semimajor axis. The initial positions for the embryos around a $1$~M$_\odot$ are 0.67,  1.00, 1.47, 2.17, 3.19, 4.69, 6.92, 10.18, 15.00 and 22.08~au, while for the non-solar cases, the corresponding initial semimajor axis scales with the disk's inner edge as $a_{\mathrm{ini}} \propto (M_\star/$M$_{\odot})^{1/3}$ following \citet{Venturini2020Letter}. This initial semimajor axes distribution ensures that, across most stellar masses, five embryos are initially placed inside the ice line and five beyond it. Finally, for each set of disk parameters we considered three different starting times for computing planetary growth: $t_{\text{start}}=0, 0.1$, and $0.5$~Myr.

Exploring all combinations of stellar and disk parameters, along with the different planet formation starting times, we computed $5400$ simulations ($900$ planetary embryos per stellar mass).
This extensive grid is sufficient to derive robust global characteristics of the emerging exoplanet populations around intermediate-mass stars.

\section{Disk evolution and planet formation}

We first illustrate gas and dust disk evolution alongside planet formation using a nominal parameter set: high disk mass (10\% of stellar mass), high viscosity ($\alpha_0=10^{-3}$), solar metallicity ($[\text{Fe}/\text{H}]=0.0$), and $t_{\text{start}}=0$.  We note that the described results are exemplary for planet formation around intermediate mass stars, but at the same time, depend on this particular set of parameters. 
General trends and the dependence of our predictions on these parameters are described in Sections~\ref{sec:global} and \ref{sec:impact_param}.  

\begin{figure*}[ht!]
  \centering
    \includegraphics[angle=0, width= \textwidth]{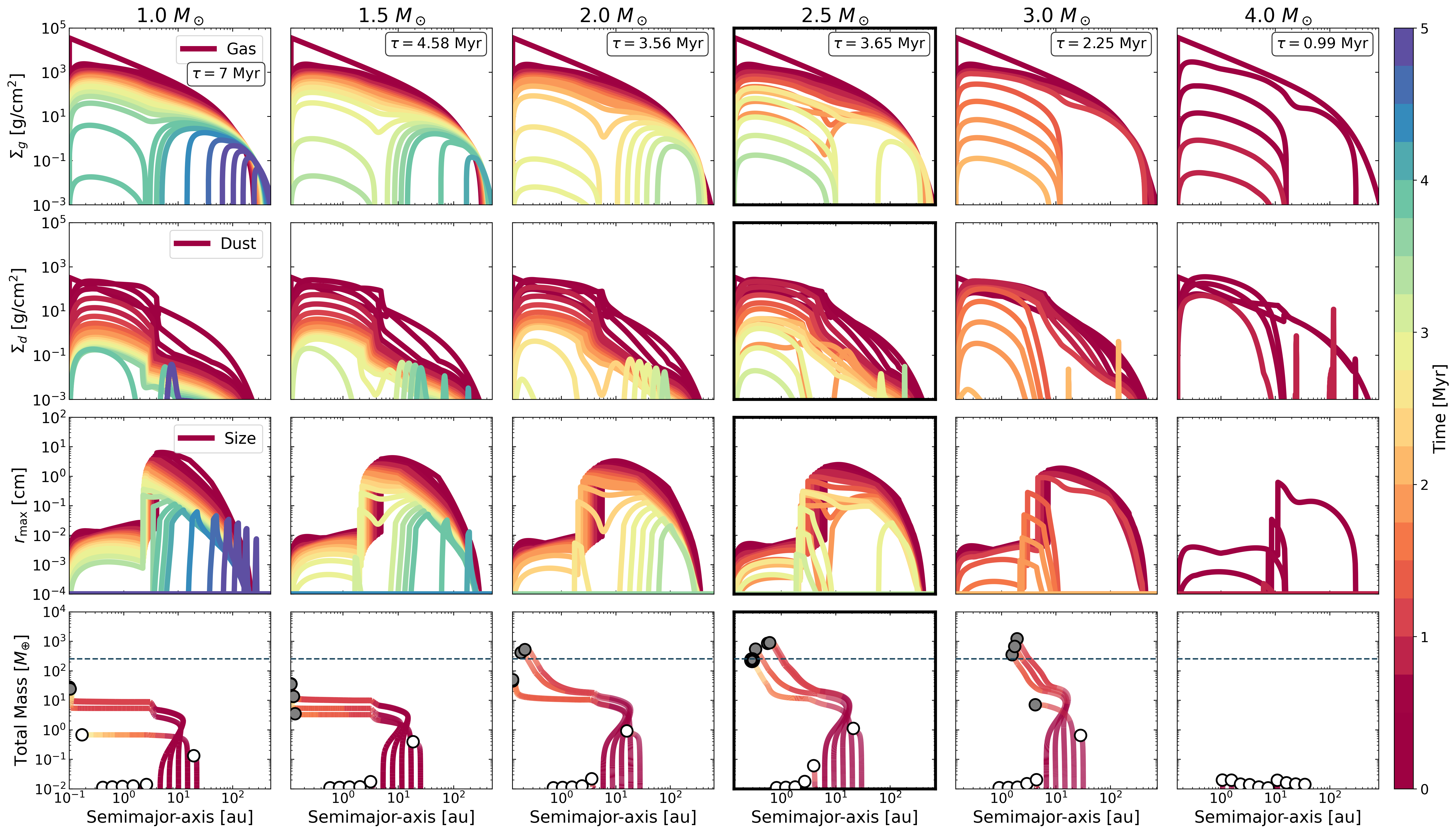}
    \caption{Disk evolution and planet formation tracks for our nominal cases: high mass disks (10$\%$), high-viscosity disks ($\alpha_0=10^{-3}$), $[\text{Fe}/\text{H}]=0.0$, and planet formation starting time $t_{\text{start}}=0$Myr. 
     From top to bottom, the rows display the time evolution of the gas surface density with a label representing the dissipation timescale, dust surface density, and maximum pebble size, followed by the planet formation tracks in the bottom row. Final planet masses and locations are marked by circles: grey for those that reached the pebble isolation mass, and white for those that did not. The subplots for the $2.5\mathrm{M}_\odot$ case, highlighted with bold borders, represent the disk evolution and planetary formation for the revenant disk case (see sect. \ref{revenant}). 
     }
     \label{Nominal-Case}
\end{figure*}

\subsection{Gas and dust evolution characteristics}

Figure~\ref{Nominal-Case} shows the time evolution of the gas (first row) and dust surface density (second row), the maximum pebble size (third row), and the planet formation tracks for each stellar mass. We refer readers to \citetalias{Ronco2024} for a detailed description of the gas disk evolution but here we summarize the most relevant aspects. 

The evolutionary pathway of gas disk dispersal is not universal, but strongly depends on the stellar mass, the initial disk mass, and the viscosity parameter. While disks around lower-mass stars typically follow the classical inside-out clearing (the cases for $1$~M$_\odot$ and $1.5$~M$_\odot$ in the top row), increasing the stellar mass shifts the evolution towards a more homogeneous dispersal (the case of $2$~M$_\odot$), where the inner and outer disks formed after gap opening vanish on similar timescales. For the most massive stars in our grid 
($3-4$~M$_{\odot}$), strong FUV irradiation drives an outside-in dispersal, where the outer disk dissipates significantly faster than the inner component.

Furthermore, in \citetalias{Ronco2024} we identified a novel evolutionary pathway, termed revenant disk evolution, which occurs for a stellar mass of $2.5$~M$_\odot$ in our nominal case. During revenant disk evolution an initial gap opened by X-ray photoevaporation reconnects due to viscous evolution as the X-ray emission drops, before a second gap is eventually opened by FUV radiation. Overall, while gas dissipation timescales generally decrease with increasing stellar mass, configurations such as the revenant phase can effectively slightly extend the gas disk's lifetime. 
Specifically, for our nominal models, the disk dissipation timescales are 7.00, 4.58, 3.56, 3.65, 2.25, and 0.99\,Myr for stellar masses of 1.0, 1.5, 2.0, 2.5, 3.0, and 4.0\,M$_\oplus$, respectively.

As shown in the second row of Fig.~\ref{Nominal-Case}, the dust surface density evolves in tandem with the gas disk. Consequently, it displays a similar transition from an inside-out dispersal around low-mass stars to an outside-in clearing for higher-mass stars. The evolution of the maximum pebble size (third row) follows the exact same pattern. A striking feature, however, is that the maximum pebble size consistently decreases towards higher stellar masses. This trend arises because both the viscosity parameter, $\alpha$, and the sound speed, $c^2_{\text{s}}$ (which scales with the disk temperature), increase with stellar mass. As these quantities govern the maximum pebble size according to Eqs. \ref{eq:drift} and \ref{eq:frag}, their increase limits the size that pebbles can reach.

\subsection{Planet formation tracks}

We find that planet formation is affected  by the different evolutionary pathways. The bottom row of Fig.~\ref{Nominal-Case} shows the time evolution of the total mass of each of the $10$ simulated planets as a function of their location on the disk. Across all stellar masses, the five innermost planets (initially located inside the water ice line) experience negligible growth, failing to even reach a Mars-mass. This lack of growth is a consequence of the high disk viscosity combined with the fragmentation threshold velocity of 1 m s$^{-1}$ adopted for silicate pebbles inside the ice line. Consequently, pebble growth in this inner region is severely limited by fragmentation, leading to pebble sizes of $\lesssim 10^{-2}$~cm (see Fig.~\ref{Nominal-Case}) with very low Stokes numbers.

In contrast, the five planets initially located beyond the ice line successfully grow beyond the mass of Mars, with the sole exception of the $4$~M$_\odot$ case, which does not form massive planets. These outer planets grow while migrating through the disk. Initially, most of them undergo outward migration driven by thermal torques, which allows them to sweep through a larger region of the disk and accrete more pebbles. Once the thermal torque becomes inefficient, they reverse their direction and begin to migrate inward driven by Lindblad torques. Eventually, they reach their pebble isolation mass (a stage easily identified in certain tracks when the planet's mass transitions from active growth to a constant value); the final mass and location of these planets are indicated by grey circles, whereas white circles denote the final state of planets that never reached this threshold.

Reaching the pebble isolation mass does not guarantee that a planet will accrete enough gas to become a gas giant. To facilitate future comparisons with observations, we define a gas giant planet as having a mass greater than $0.8~$M$_\text{J}$ following \citet{Wolthoff2022} (see the dashed blue line in the bottom row of Fig. \ref{Nominal-Case}). Under this criterion, our simulations show that $1$, $1.5$, and $4$~M$_\odot$ stars do not produce gas giants. In contrast, systems around $2$, $2.5$,  and $3$~M$_\odot$ stars successfully form them.

The inability to form gas giants around the lower-mass stars ($1$~M$_\odot$ and $1.5$~M$_\odot$) is a direct consequence of their rapid orbital decay. In these systems, planets that reach the pebble isolation mass migrate inward too fast to accrete gas significantly. This is not the case for the $2$, $2.5$, and $3$~M$_\odot$ systems, where gas giant formation is possible. Because migration rates are inversely proportional to stellar mass \citep{Tanaka02}, planets around massive stars migrate inward more slowly. This gives them enough time to undergo runaway gas accretion.

Finally, the evolutionary pathway of the disk and its dissipation timescale also play a crucial role in shaping planet growth and migration. In low-mass stellar systems, the disk dissipation timescales are extended, and the photoevaporation gap opens relatively late (between $3$ and $4$\,Myr). By this time, the planets have already formed and migrated toward the inner regions of the disk. In contrast, for higher stellar masses (such as $3$ and $4$~M$_\odot$), the dissipation timescales are significantly shorter. Driven by the outside-in dispersal, planets forming in the outer regions face two possible outcomes: either they manage to grow into gas giants but fail to migrate inward because they become trapped in low gas-density environments (beyond the gap location opened by FUV photoevaporation), or they simply do not have enough time to grow because the local gas dissipates extremely rapidly. 

\subsection{Planet formation in revenant disks}
\label{revenant} 
Our nominal case for the $2.5$~M$_\odot$ star presents a revenant disk evolution \citepalias[described in detail in Sect. 3.4 of][]{Ronco2024}.In these scenarios, the dust surface density closely follows the evolution of the gas. Specifically, when photoevaporation opens a gap in the gas disk, it simultaneously clears a gap in the solid component. As the gas disk is subsequently rebuilt due to the decay in X-ray photoevaporation, the dust disk follows the same behavior: the inner and outer disks reconnect, allowing solid material to flow into the inner regions once again \citepalias{Ronco2024}. These dynamics can significantly influence planet formation and contribute to the growth of more massive planets.

To fully understand this, it is necessary to examine simultaneously and in detail both the disk profile evolution and the planet formation track of the penultimate planet (the 9th planet in terms of initial orbital distance, highlighted with a thicker black circle in fig. \ref{Nominal-Case}) for the $2.5$~M$_\odot$ case. This planet reaches its pebble isolation mass ($\sim11$~M$_\oplus$) at $\sim1.44$~Myr while located at $\sim$3.18\,au. Subsequently, by 2 Myr, the planet migrates inward to approximately 1 au, doubling its mass through gas accretion without gaining any additional solid material. At this stage, a gap opens in both the gas and dust disks, depleting the available gas in the inner region and halting the influx of pebbles from the outer disk. While the gap remains open, the planet reaches a mass of $\sim 50$~M$_\oplus$ by accreting the residual gas from the inner disk. However, this gap is short-lived; as X-ray photoevaporation becomes inefficient, the disk replenishes, restoring the inward flux of both gas and solids. During this second evolutionary stage, the renewed gas supply allows the planet to resume its growth, ultimately reaching a final mass of $\sim 228$~M$_\oplus$.

\section{Global results and general trends}
\label{sec:global}

\begin{figure*}[ht!]
  \centering
    \includegraphics[angle=0, width= \textwidth]{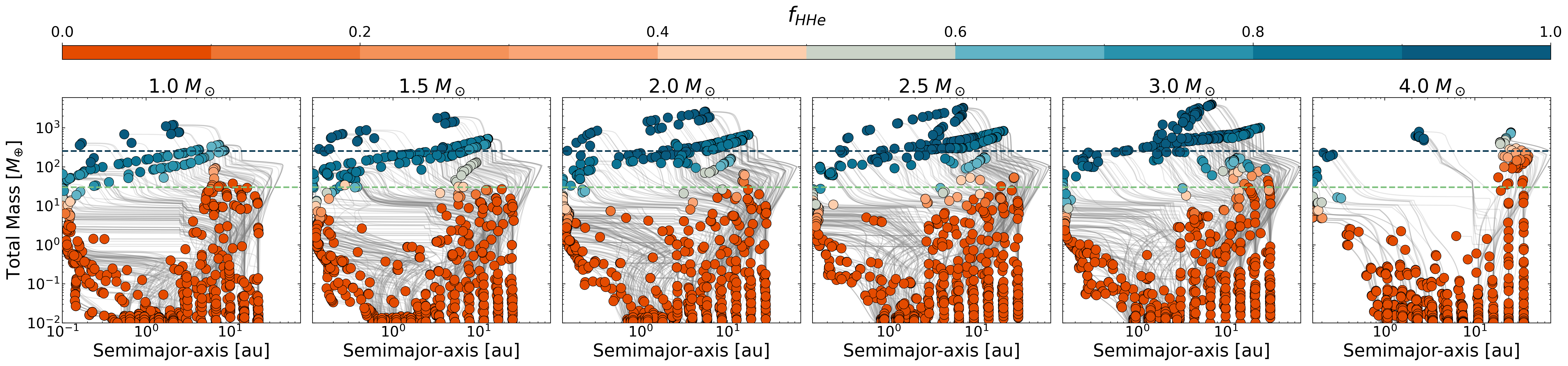}
    \caption{Planet formation tracks (grey curves) and final planet total mass and location (colored circles) computed for each stellar mass. The color palette represents the planet H-He mass fraction at the end of its formation. The blue and green dashed horizontal curves denote the lower limits of the gas giants ($M_p > 0.8$~M$_{\mathrm{J}}$) and intermediate-mass planets ($30$~M$_\oplus < M_p < 0.8$~M$_{\mathrm{J}}$).}
\label{Planet-Formation-Tracks}
\end{figure*}

\begin{figure*}[ht!]
  \centering
    \includegraphics[angle=0, width= \textwidth]{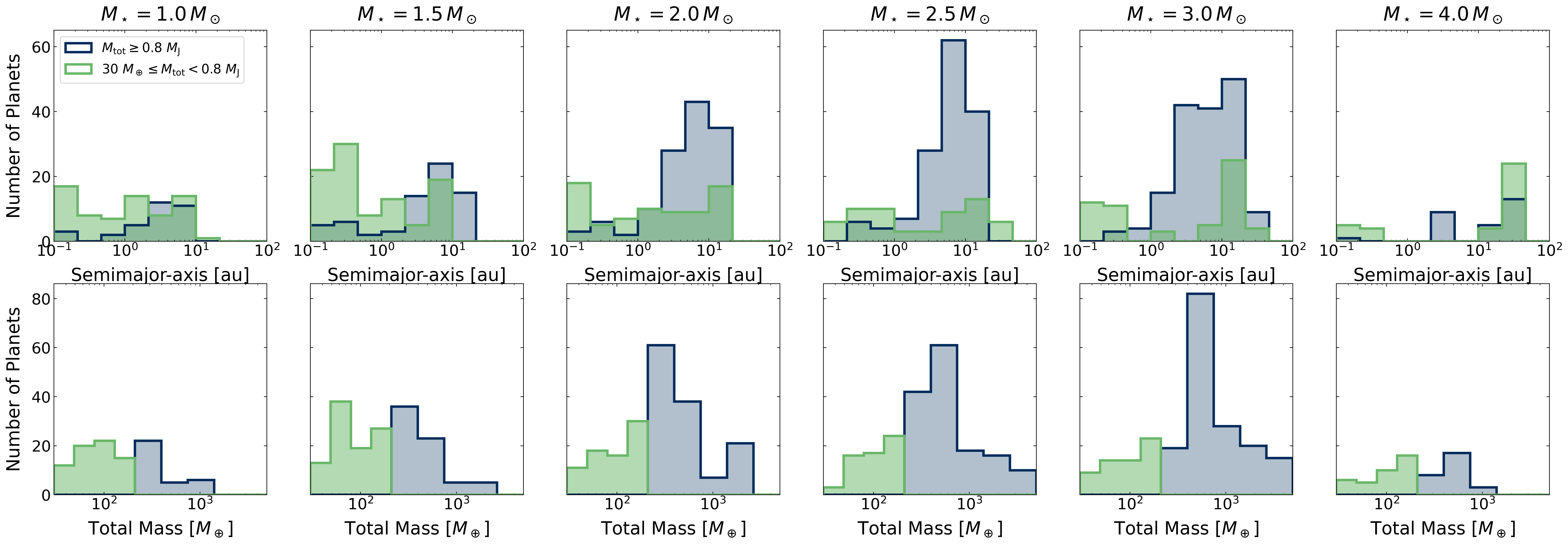}
    \caption{Final semimajor axis (top row) and mass distributions (bottom row) of the simulated planets, separated by stellar mass. The populations are divided into intermediate-mass planets ($30$~M$_{\oplus} < M_p \leq 0.8$~M$_{\text{J}}$) and gas giants ($M_p > 0.8$M$_{\text{J}}$).} 
\label{Histograms-Semimajoraxis-Masses}
\end{figure*}

Here, we present the overall results obtained from the complete set of simulations. Figure~\ref{Planet-Formation-Tracks} shows all planet-formation tracks and their final masses and orbital locations for each stellar mass considered. The colour scale indicates the final H–He mass fraction, while the green and blue horizontal dashed lines mark the upper mass boundaries adopted for low-mass ($30$\,M$_\oplus$) and intermediate-mass planets ($0.8$\,M$_{\rm J}$), respectively. A clear systematic trend emerges: both the number of giant planets and their characteristic masses increase with stellar mass up to $3$\,M$_\odot$. This trend breaks down abruptly at $4$\,M$_\odot$, around which only very few gas giants form. Notably, substantial populations of lower-mass planets are produced for all stellar masses considered.

Figure \ref{Planet-Formation-Tracks} also reveals a distinct region in the mass-semimajor axis diagram, spanning roughly from $0.2$ to $2$~au and masses between $0.1$~M$_\oplus$ and $\sim 30$~M$_\oplus$, that exhibits a clear deficit of planets across all scenarios, resulting from the interplay of several physical processes. This depleted region is bounded at the top (above $\sim 30$~M$_\oplus$) by the intermediate-mass planets, which more or less continuously populate the entire range of semimajor axes. To the left (inside $\sim 0.2$~au), it is constrained by planets that, whether affected by thermal torques or not, managed to migrate rapidly toward the inner edge of the disk. Finally, the right boundary (outward of $\sim 2$~au) is defined by a population of planets with masses between $0.1$~M$_\oplus$ and $30$~M$_\oplus$ that began forming beyond the ice line but were neither able to grow enough to become gas giants nor migrated inward.

 The top row of Fig.~\ref{Histograms-Semimajoraxis-Masses} shows the semimajor axis distribution for each scenario, highlighting planets with masses between $30$~M$_\oplus$ and $0.8$~M$_\text{J}$ (green) and giant planets with masses above $0.8$~M$_\text{J}$ (blue). This plot not only clearly illustrates the increasing number of giant planets with stellar mass, but also shows that, as a consequence of thermal torques in combination with the adopted Type II migration rates from \citet{Kanagawa2018} and the disk evolution pathways, these giants are mainly located beyond a few au 
 \citep[similar results were found by][employing the same type II migration rates]{Ida2018}, with their semimajor axes increasing for higher stellar masses. 
In Appendix~\ref{Ap:TT} we discuss the implications of considering thermal torques. 
For the $4$~M$_\odot$ case, although the number of giant planets drops abruptly as previously mentioned, the few giants that do form are still preferentially located far from the central star. Regarding planets with masses between $30$~M$_\oplus$ and $0.8$~M$_{\mathrm{J}}$, there is no apparent preferred orbital location or clear trend with stellar mass. However, an enhancement of these objects is noticeable very close to the host star as well as in regions beyond a few au.
The bottom panel of Figure \ref{Histograms-Semimajoraxis-Masses} displays the planetary mass distributions. While giant planet masses clearly increase with stellar mass (see Fig.~\ref{Planet-Formation-Tracks}), they never exceed $\sim 3$~M$_\text{J}$, with the $4$~M$_\odot$ case failing to form planets above $2.5$~M$_\text{J}$.

\section{Impact of disk parameters and planet formation starting time on occurrence rates}
\label{sec:impact_param} 

The occurrence rate of a given planet class can be calculated as the fraction of simulations successfully producing a planet of that class. For the gas giants considered here, we define
\begin{equation}
    f_{\rm giant} =
    \frac{N(M_{\rm p}>0.8\,M_{\rm J})}{N_{\rm sim}},
\end{equation}
where $N_{\rm sim}$ is the total number of simulated embryos in the corresponding subgroup. The uncertainties are estimated using the standard error of a binomial proportion, that is, $\sigma_f =\sqrt{f_{\rm giant}(1-f_{\rm giant})/{N_{\rm sim}}}$. 

The interpretation of this fraction as a stellar planet occurrence rate depends on the adopted initial conditions. Population synthesis studies commonly compare synthetic planet frequencies with observations while prescribing the initial number, masses, locations, and formation times of planetary embryos \citep{emsenhuberetal21-1,Mulders2021, Pan2025, johnston_2026}.   
These prescriptions are not uniquely constrained by observations and constitute additional assumptions underlying the predicted occurrence rates. 

In our calculations, each simulation follows a single embryo, and the ten initial orbital locations represent independent realizations rather than embryos growing simultaneously within the same disk. Moreover, the explored disk parameters and embryo starting times are sampled discretely and given equal weight, rather than drawn from empirically constrained distributions. Our calculated fractions therefore quantify the efficiency of giant planet formation within the adopted parameter grid; their identification with the fraction of stars hosting giant planets is conditional on these assumptions. 

\begin{figure*}[ht!]
  \centering
    \includegraphics[angle=0, width= 0.95\textwidth]{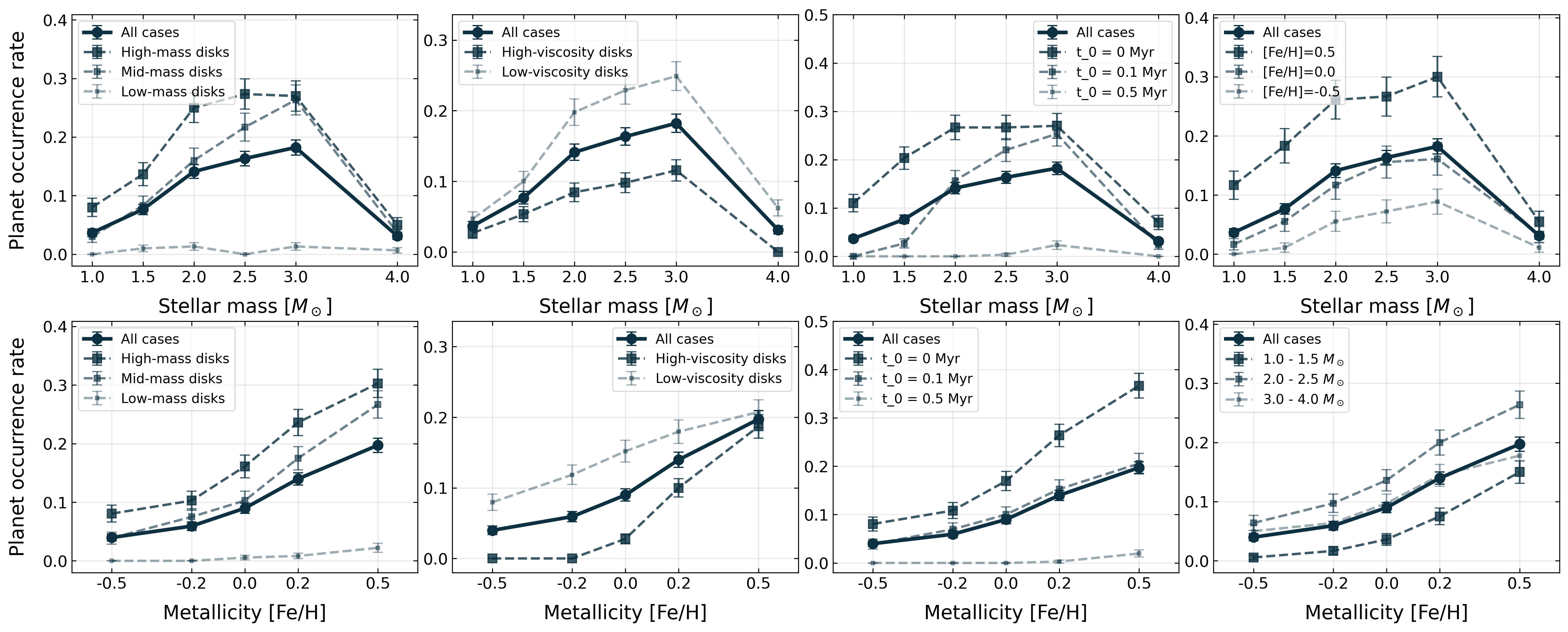}  
    \caption{Occurrence rates of giant planets $M_p > 0.8\ M_J$) as a function of stellar mass (top row) and metallicity (bottom row). Solid blue lines with circles show the global average across all parameters. Dashed lines with squares isolate specific initial conditions: disk mass (column 1), viscosity $\alpha$ (column 2), planet formation starting time (column 3), and metallicity/stellar mass bins (column 4). Error bars denote 1$\sigma$ binomial confidence intervals.}
\label{Planet-Occurence-Rates}
\end{figure*}

Figure~\ref{Planet-Occurence-Rates} shows the resulting gas-giant occurrence rates as a function of stellar mass in the top row and metallicity in the bottom row. The four columns illustrate the sensitivity of the results to the adopted disk mass, viscosity, planet-formation starting time, and metallicity, respectively.

The top row reveals a robust qualitative dependence on stellar mass. For all model variations, the occurrence rate of gas giants increases with stellar mass up to $2.5$--$3\,$M$_\odot$, but drops abruptly between $3$ and $4\,$M$_\odot$. The precise location and height of the maximum nevertheless might depend on the adopted disk properties and, as we will discuss in Appendix \ref{Ap:PhotoRates}, on the adopted photoevaporation rates.

Disk mass has the strongest effect on the absolute occurrence rate (first column). High-mass disks produce substantially more gas giants, with the occurrence rate reaching its maximum around $2.5\,$M$_\odot$, whereas low-mass disks almost completely fail to form giant planets.
The occurrence rate also depends strongly on viscosity (second column). Models with $\alpha_0=10^{-4}$ produce more gas giants than those with $\alpha_0=10^{-3}$. Lower turbulent viscosity allows particles to grow to larger sizes and increases the efficiency of pebble accretion, thereby facilitating the rapid formation of sufficiently massive cores that can subsequently accrete substantial gaseous envelopes. 

The starting time of planet formation in our models has an impact on the occurrence rate but also the dependence on stellar mass (third upper panel). If planet formation starts at $0.5$~Myr, very few planets are formed. The earlier planet formation sets in, the more planets are formed. If planets form right from the start of disk evolution the maximum occurrence rate corresponds to a broad plateau from $2$-$3$~\Msun while it otherwise peaks at $3$~\Msun. 

The stellar-mass dependence is recovered at each metallicity considered: the occurrence rate generally rises towards $2$--$3\,\mathrm{M}_\odot$ and subsequently decreases sharply at $4\,\mathrm{M}_\odot$. For the two extreme metallicities, $[\mathrm{Fe/H}]=\pm0.5$, the maximum occurs at $3\,\mathrm{M}_\odot$, whereas at solar metallicity it is located near $2.5\,\mathrm{M}_\odot$ (fourth upper panel of Fig.\,\ref{Planet-Occurence-Rates}).

Part of the increase in giant-planet occurrence towards higher stellar masses may reflect the adopted initial conditions. In our models, disk mass and viscosity vary with stellar mass, and, even at fixed metallicity, the dust-to-gas ratio increases with stellar mass. However, the abrupt decline between $3$ and $4\,\mathrm{M}_\odot$ cannot be produced by these initial scalings, which would instead favor a continued increase. The downturn therefore represents a genuine consequence of the more rapid evolution and dispersal of disks around $4\,\mathrm{M}_\odot$ stars, which limits the time available for planetary cores to grow and accrete massive gaseous envelopes. 

The bottom row shows the dependence of the gas-giant occurrence rate on metallicity. As expected, metal-rich disks produce more giant planets because their larger solid content promotes the formation of massive cores. 
This increase is strongest for high-mass disks, low viscosities, early onset of planet formation, but nearly independent of stellar mass. 

\section{Comparisons with Observations}

In this section we aim to compare our simulation results with the observational data from \citet{Wolthoff2022}. We approach this comparison as carefully as possible, acknowledging the inherent limitations of our work: our study does not constitute a full population synthesis model, nor was its primary goal to reproduce the observed occurrence rates of giant planets. Comparisons with observed occurrence rates should consequently be interpreted as an exploration of trends and parameter sensitivity, rather than as a direct determination of absolute stellar hosting frequencies.

\subsection{Global giant planet occurrence rates}
\label{sec:Pesado}

\begin{figure}[ht!]
  \centering
    \includegraphics[angle=0, width= 0.9\columnwidth]{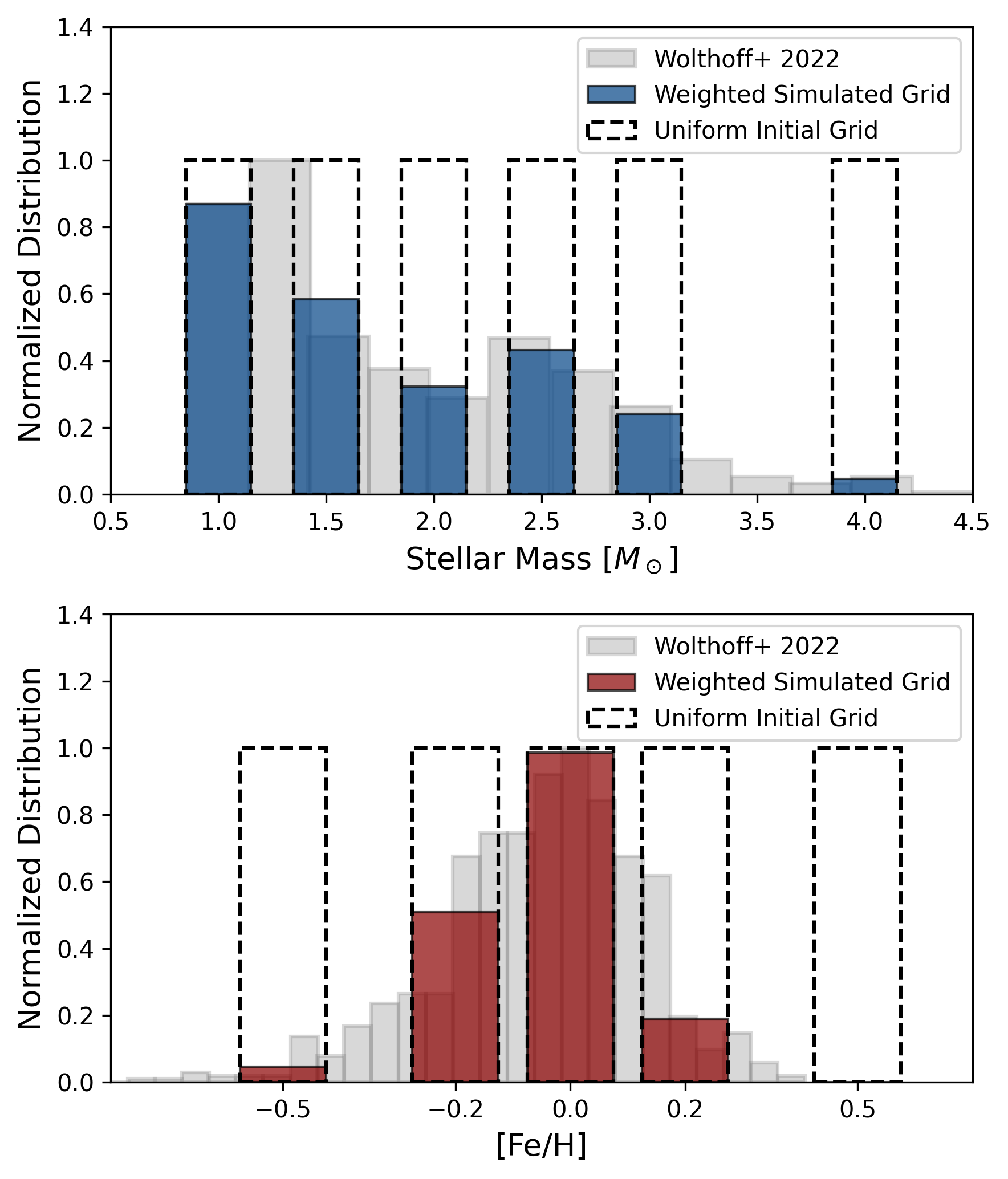}   
    \caption{Comparison between the stellar sample of 482 stars from \citet{Wolthoff2022} (grey histograms) and the discrete parameter space adopted for our planetary formation simulations represented with the black dashed bars in both plots. \textit{Top panel:} Normalized stellar mass distribution. The blue bars represent the statistical weights assigned to each simulated stellar mass ($1.0, 1.5, 2.0, 2.5, 3.0$, and $4.0 \, \text{M}_\odot$). \textit{Bottom panel:} Normalized metallicity distribution. The red bars indicate the corresponding weights for the simulated [Fe/H] values ($-0.5, -0.2, 0.0, 0.2$, and $0.5$ dex).} 
\label{fig:wolthoff_distributions}
\end{figure}

Our theoretical framework relies on a discrete and uniform grid of stellar masses (1.0, 1.5, 2.0, 2.5, 3.0, 4.0 $\mathrm{M}_\odot$) and metallicities (-0.5, -0.2, 0.0, 0.2, 0.5 dex). However, the observational sample of 482 stars in \citet{Wolthoff2022} is not uniformly distributed across this parameter space. Thus, to ensure a fair comparison with the results of \citet{Wolthoff2022}, we must correct for the demographic biases inherent in their sample. To do so we implemented a weighting scheme based on the mass and metallicity distributions of their sample. For each node $(M_\star, \mathrm{[Fe/H]})$ in our simulated grid, each simulated system $i$ is assigned a statistical weight, $w_i= w_{M_\star}(M_{\star, i}) \times w_{\rm [Fe/H]}(\mathrm{[Fe/H]}_i)$. The factors $w_{M_\star}$ and $w_{\rm [Fe/H]}$ are derived from the normalized stellar mass and metallicity distributions of \citet{Wolthoff2022} (see their Figure 1). We calculate these weights by interpolating the observed distributions at the exact values of the simulated parameters (as illustrated in the blue and red histograms in Figure \ref{fig:wolthoff_distributions}).

By applying these weights, we transform our homogeneous theoretical grid into a population that mimics the demographic properties of the observational stellar sample. This allows us to compute a weighted global occurrence rate for gas giants that can be fairly compared to the $10.7^{+2.2}_{-1.6}\,\%$ rate reported by \citet{Wolthoff2022}.

The weighted occurrence rate, $F_{\text{w}}$, is given by:
\begin{equation}
    F_{\text{w}} = \frac{\sum_{i} \delta_i \, w_i}{N_{\rm sys}} \times 100 \, \%,
    \label{eq:occurrence_rate}
\end{equation}
where $\delta_i$ is a boolean selection function equal to 1 if the simulation produces a gas giant fulfilling the \citet{Wolthoff2022} mass and period criteria ($M_p > 0.8 M_{\text{J}}$, and the period between 89 and 3600 days), and 0 otherwise. Given that we compute 180 models for each $(M_\star, \mathrm{[Fe/H]})$ combination, the weighted number of host systems, $N_{\rm sys}$, is given by:
\begin{equation}
    N_{\rm sys} = \sum_{m \in M_\star} \sum_{f \in \mathrm{[Fe/H]}} 180 \cdot w_{M_\star}(m) \cdot w_{\rm [Fe/H]}(f).
    \label{eq:nsys}
\end{equation}
Without filtering by orbital period, our weighted global occurrence rate reaches 7.43\%. This value represents the total intrinsic formation and survival efficiency of our models across all orbital separations, rather than an observable frequency. However, when restricting our synthetic sample to orbital periods between 89 and 3600 as in \citet{Wolthoff2022}, our weighted rate decreases to 2.83\%, well below their reported 10.7\%. This substantial drop highlights a key physical outcome of our setup: the inclusion of thermal torques drives strong outward migration, effectively parking a significant fraction of gas giants beyond the current radial velocity detection limits. If these thermal torques are less efficient in reality, or if our current implementation could be overestimating the extent of this outward migration, as discussed in Appendix \ref{Ap:TT}, a larger fraction of these giants would remain within the intermediate-period window. This would naturally bring our filtered occurrence rates closer to the $10.7\%$ observed by \citet{Wolthoff2022} and similar radial velocity surveys \citep{Wittenmyer2020}.

\subsection{Planet occurrence rate as a function of period}
\label{sec:period}

\begin{figure}[ht!]
  \centering
    \includegraphics[angle=0, width= 0.9\columnwidth]{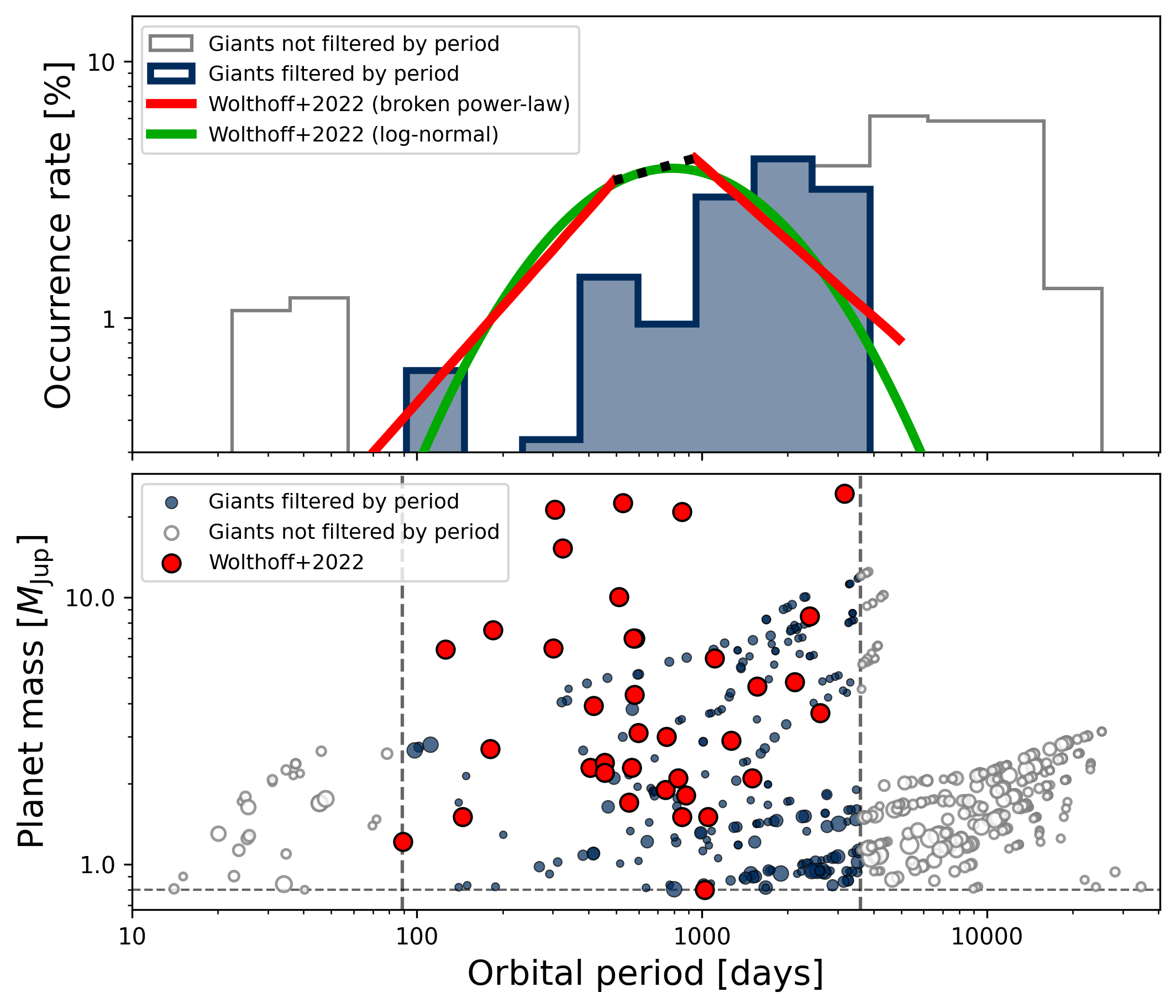}   
    \caption{\textit{Bottom panel:} Simulated giant planets ($M_{\rm p} > 0.8 \, M_{\rm J}$) in the mass-period plane. The marker size of the simulated planets is proportional to their statistical weight, derived from the stellar mass and metallicity distributions. Blue filled circles highlight the planets located within the observational window (periods between 89 and 3600 days, represented by vertical dashed lines), while white hollow circles represent planets outside this range. Red points show the exoplanet sample from \citet{Wolthoff2022}. \textit{Top panel:} Planet occurrence rate as a function of orbital period. The grey step-histogram shows the unfiltered simulated giants, whereas the blue filled histogram represents the occurrence strictly within the observational period limits. For comparison, the empirical broken power-law and log-normal fits from \citet{Wolthoff2022} (see their fig. 8) are overlaid in red and green, respectively.}
\label{fig:Ajuste-Periodo}
\end{figure}

We present in Figure \ref{fig:Ajuste-Periodo} the mass-period distribution of our simulated giant planets ($M_{\rm p} > 0.8 \, M_{\rm J}$) alongside their corresponding weighted occurrence rates. In the bottom panel, we simultaneously show the minimum planet mass of the exoplanet sample of \citet{Wolthoff2022} in red circles (see their fig. 7) together with our synthetic planet population. The size of each simulated planet is scaled proportionally to its statistical weight, which is derived from the observational stellar mass and metallicity distributions (Section \ref{sec:Pesado}). We highlight planets within the \citet{Wolthoff2022} detection window (89–3600 days) in blue, leaving those outside as hollow circles.

The top panel of Figure \ref{fig:Ajuste-Periodo} shows histograms of the occurrence rate as a function of the orbital period. The unfiltered simulated population (grey step-histogram) shows a broad distribution. However, when we apply the strict observational period filter (blue filled histogram), the theoretical occurrence rate drops, enabling a direct and fair comparison with the broken power-law and log-normal fits derived by \citet{Wolthoff2022}. 

As shown in the bottom panel, there is a clear overlap between our synthetic exoplanet population and the observational sample from \citet{Wolthoff2022}. Nevertheless, our calculated occurrence rates appear shifted toward longer orbital periods, falling short of the values predicted by their log-normal and broken power-law fits.

This shift is a clear consequence of the combined effects of the adopted migration rates and thermal torques, which might be overestimated (Appendix \ref{Ap:TT}). Nevertheless, it is interesting to highlight that our simulations do not produce planets more massive than $\sim10 M_{\text{J}}$ in agreement with \citet{Wolthoff2022}. Furthermore, in both their observational sample and our synthetic one, the vast majority of gas giants have minimum masses below $3 M_{\text{J}}$. 

\subsection{Planet occurrence rates as a function of stellar mass and metallicity}

\begin{figure*}[ht!]
  \centering
    \includegraphics[angle=0, width= 0.95\textwidth]{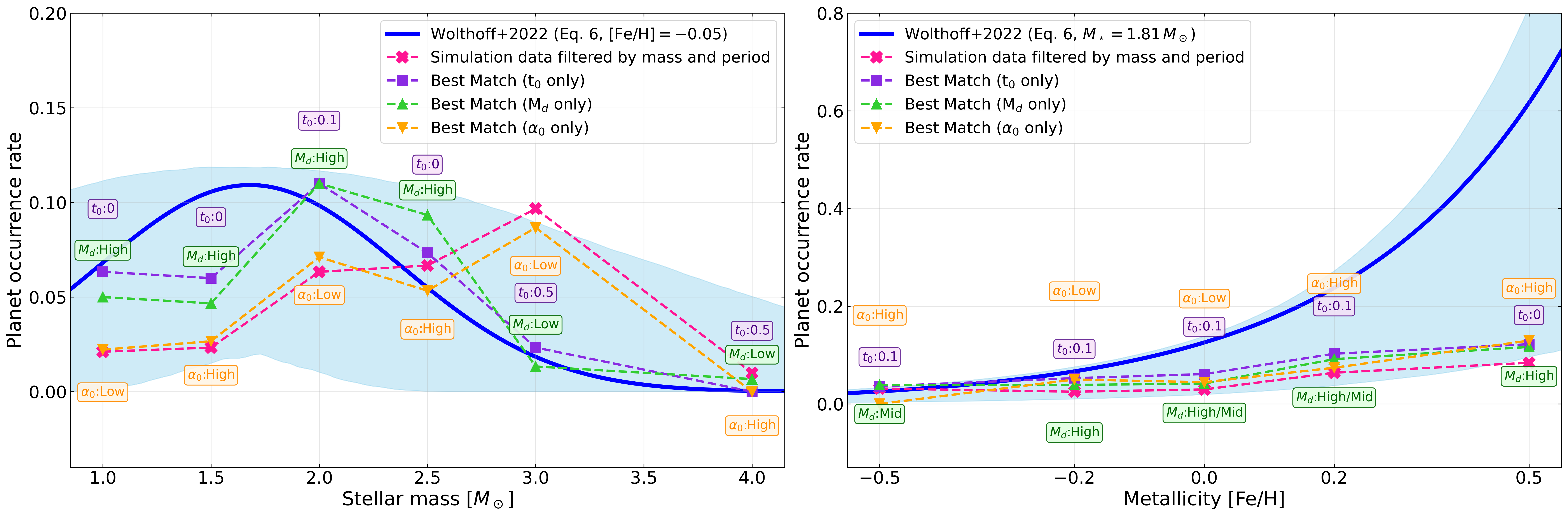}   
    \caption{Theoretical occurrence rates from our planetary formation simulations compared against the observational trends from Fig. 11 in \citet{Wolthoff2022}. \textit{Left panel:} Planet occurrence rate as a function of stellar mass. \textit{Right panel:} Planet occurrence rate as a function of stellar metallicity. The solid blue lines represent the observational fitted occurrence of \citet{Wolthoff2022} as a function of stellar mass (left panel) with the metallicity set to their sample mean of $[$Fe/H$] = -0.05$ and as a function of metallicity (right panel) with the mass set to their sample mean of $M_\star= 1.81\mathrm{M}_\odot$. The shaded regions represent the $1\sigma$ uncertainty in eq. \ref{eq:wolthoff_eq6} (their eq. 6). In both panels the magenta crosses connected by dashed lines indicate the occurrence rate derived from our entire simulated grid, filtered by giants ($M_p > 0.8M_J$ and $89~\text{days} < P < 3600~\text{days}$). The additional curves represent the "Best Match" scenarios, where a single initial physical parameter, disk mass ($M_d$, green triangles), formation time ($t_0$, purple squares), or viscosity ($\alpha_0$, orange circles), is isolated to minimize the discrepancy with the  observations. The floating labels adjacent to these optimal points explicitly denote the specific parameter value required to best reproduce the observational constraints at each mass and metallicity node.}
\label{fig:Mejor-Ajuste}
\end{figure*}

We now compare our synthetic population with the giant planet occurrence rates derived by \citet{Wolthoff2022}, who parameterized it through a combined Gaussian mass dependence and a power-law metallicity given by:
\begin{equation}
    f(M_\star, \mathrm{[Fe/H]}) = C \cdot \exp\left[ -\frac{1}{2} \left( \frac{M_\star - \mu}{\sigma} \right)^2 \right] \cdot 10^{\beta \, \mathrm{[Fe/H]}},
    \label{eq:wolthoff_eq6}
\end{equation}
where the best-fit parameters obtained from their Markov Chain Monte Carlo (MCMC) analysis are $C = 0.128 \pm 0.027$, $\mu = 1.68 \pm 0.59 \, \mathrm{M}_\odot$, $\sigma = 0.70 \pm 1.02 \, \mathrm{M}_\odot$ and $\beta = 1.38 \pm 0.36$.

The solid blue lines in Figure \ref{fig:Mejor-Ajuste} represent the occurrence rate given by equation~(\ref{eq:wolthoff_eq6}). In the left panel, it is evaluated as a function of stellar mass by setting the metallicity to the sample average, $\langle \mathrm{[Fe/H]} \rangle = -0.05$\,dex. In the right panel, it is evaluated as a function of metallicity for the average stellar mass of the survey, $\langle M_\star \rangle = 1.81 \, \mathrm{M}_\odot$. The light blue shaded band illustrates the $1\sigma$ uncertainties.

Following \citet{Wolthoff2022}, we again restrict our synthetic exoplanet sample to those with masses of $M_p > 0.8 \, M_J$ and orbital periods between 89 and 3600 days. It is important to mention that this period restriction decreases our total giant planet sample by more than 50\%. The occurrence rate computed over the full, unweighted simulation suite ($N = 900$ systems per stellar mass bin in the left panel, and $N = 1080$ systems per metallicity bin in the right panel) is represented by the pink dashed curve with cross markers. Once again we see that 
while the overall synthetic pink curve captures the rise toward intermediate masses ($2.0$-$3.0\,\mathrm{M}_\odot$), our occurrence rates up to $2\mathrm{M}_\odot$ underpredict the observational rates, and beyond $2\mathrm{M}_\odot$ overpredict them, still finding the peak at around $3\mathrm{M}_\odot$. Also, while we predict a correlation with metallicity (Fig.~\ref{fig:Mejor-Ajuste}, right panel) that agrees with the one derived from observations taking into account $1\sigma$ uncertainties of the latter, the global average lies systematically below the blue curve. 

To investigate whether particular initial conditions provide better agreement with the occurrence rates derived by \citet{Wolthoff2022}, we examined the dependence of our results on three key parameters: the initial disk mass $M_d$, the onset of planet formation $t_0$, and the viscosity parameter $\alpha_0$. For each stellar mass, we divided the simulations into subsets according to the value of one of these parameters, keeping all other parameters unrestricted. We then calculated the giant planet occurrence rate for each subset and identified the parameter value that most closely reproduces the observational prediction. The resulting curves, shown in Fig.~\ref{fig:Mejor-Ajuste}, illustrate how the agreement with observations changes when different initial conditions are selected at each stellar mass.

Remarkably, despite the discrete parameter grid, the limited number of simulations, and the additional restriction on orbital periods, the occurrence rates obtained by selecting the most favorable disk masses (green upward-pointing triangles) reproduce the observed stellar-mass dependence reasonably well, with the notable exception of a significant underprediction at $1.5$\,M$\odot$. The preferred disk masses suggest that relatively massive disks are needed to reproduce the observed occurrence rates around stars between $1$ and $2.5$\,M$\odot$, whereas lower-mass disks provide better agreement at higher stellar masses. This trend is broadly consistent with recent observational constraints on disk masses \citep{Stapper2026}.

A similarly good agreement is obtained when selecting the most suitable mass dependence for the onset of planet formation (violet squares). In this case, the observations favor an early start of planet formation around stars with masses between $1$ and $2.5$\,M$_\odot$, while later formation times are preferred for more massive stars. Physically, delaying the onset of planet formation reduces the amount of gas and solids available for planetary growth, potentially mimicking the effect of an initially less massive disk. Although the occurrence rates obtained by selecting the most suitable mass dependence of the viscosity parameter $\alpha_0$ (yellow downward-pointing triangles) remain within the observational $1\sigma$ uncertainty region, no systematic dependence on viscosity emerges: the preferred values alternate between high- and low-viscosity disks across neighboring stellar-mass bins which lacks a physical explanation. 

For none of the tested parameters, optimizing the dependence stellar mass does not change the predicted dependence on metallicity significantly. In all cases the dependence on metallicity is weaker than the one derived from observations but remains within the $1\sigma$ uncertainty. 
 
\section{Discussion}
\label{sec:discussion}
 
In this section we compare our model and results with previous planet formation works around intermediate-mass stars, we discuss the implications of our planet population results in the context of future polluted white dwarfs, and we finally present the main caveats in our modeling and work approach.

\subsection{Comparisons with previous planet formation works}

Among recent studies of giant planet formation around intermediate-mass stars based on pebble accretion, the works of \citet{johnston2024,johnston_2025,johnston_2026} provide the closest comparison to our models. While \citet{johnston_2026} performed a population synthesis aimed at reproducing the observed stellar-mass dependence of giant planet occurrence rates, our primary goal is to investigate the planetary populations emerging from a detailed treatment of disk evolution and planet formation.

The two approaches differ in several important respects. \citet{johnston_2026} adopt time-independent X-ray photoevaporation rates, whereas we account for the evolving X-ray and FUV irradiation of the host star and the resulting changes in disk structure and dispersal pathways. As demonstrated in Appendix \ref{App:Msyst}, these differences substantially affect disk lifetimes and planet formation, particularly around the most massive stars. Furthermore, while \citet{johnston_2026} assume a fixed pebble size of 1 mm, we explicitly model dust growth and evolution. Finally, our inclusion of thermal torques, neglected in their models, favors the formation of giant planets at wider orbital separations.

These differences naturally lead to distinct predictions for the resulting planetary architectures. By constraining their initial accretion rates and disk lifetimes with observational samples, \citet{johnston_2026} successfully reproduce the steep decline in giant planet occurrence beyond $2\,\mathrm{M}_\odot$ and recover the observed peak at $\sim 1.7\,\mathrm{M}_\odot$ from \citet{Wolthoff2022}. In our framework, the high efficiency of gas and pebble accretion in massive disks pushes the theoretical peak of giant planet formation toward $\sim 3\,\mathrm{M}_\odot$ before intense FUV photoevaporation abruptly stops it. However, as we discuss in Appendix \ref{Ap:PhotoRates}, our models show that adopting a stronger X-ray-driven mass-loss rates, such as those by \citet{Owen2012} used in the \citet{johnston_2026} model, rapidly truncates disk lifetimes around more massive stars, naturally shifting the theoretical peak of giant planet occurrence closer to $\sim 2\,\mathrm{M}_\odot$. Furthermore, the two frameworks predict fundamentally different orbital distributions. While the models of \citet{johnston_2026} allow for efficient inward migration that populates the inner regions ($\lesssim 5$ au) accessible to RV surveys, our inclusion of thermal torques combined with the disk evolutionary pathways found in \citetalias{Ronco2024} typically stalls inward migration at larger orbital radii. Consequently, our models predict a robust population of wide-orbit gas giants ($> 5$ au). As discussed in Section \ref{sec:period}, this suggests that the majority of giant planets formed around intermediate-mass stars may reside beyond the detection limits of current RV surveys, highlighting the necessity of combining RV statistics with direct-imaging surveys to capture the complete demographic picture.

\subsection{Implications for Polluted White Dwarfs}

As noted in the introduction and \citetalias{Ronco2024}, planets around intermediate-mass stars are key to understanding the atmospheric pollution and planet populations at their white dwarf remnants. Although direct detections of planets orbiting WDs are scarce \citep{Gaensicke2019,Vandenburg2020,Blackman2021,Limbach2024,Mullally2024},  roughly a third exhibit metal absorption lines \citep[e.g.,][]{Koester2014}. This contamination is attributed to the accretion of surviving planetary material that originally formed beyond a few au to withstand stellar evolution \citep[e.g.,][]{Jura2003, Gaensicke2006, MustillVillaver2012, Veras2014, Manser2019, Ronco2020, Veras2021}. Furthermore, $\sim$40\% of field WDs have masses of $0.6-0.75 M_{\odot}$ \citep{Tremblay2016}, descending directly from $1.5-3 M_{\odot}$ main-sequence progenitors \citep{Cummings2018,Marigo2020}. This implies efficient planet formation in these systems, consistent with our findings. 

Recent population synthesis models by \citet{Mauch-Soriano2026} and \citet{Miranda2026}, based on observed gas giant distributions from \citet{Wolthoff2022}, predict 
the population of substellar companions around white dwarfs. Interestingly, the likelihood determined by \citet{Mauch-Soriano2026} peaks precisely for remnants of $0.53-0.66 M_{\odot}$.
The overall predicted occurrence rate of surviving Jupiter-like planets around WDs is low ($\lesssim 3 \pm 1.5\%$). However, the \citet{Wolthoff2022} distributions are based on radial velocity surveys which might overlook 
companions at large separations \citep{Vigan_2021} and which are not sensitivve planets less massive than $0.8$\Mjup. 
These uncertainies are hard to fully solve observationally which reinforces the importance of modeling and characterizing planetary architectures around $1.5-3 M_{\odot}$ progenitors as we intend here. 

\subsection{Model and approach limitations}
\label{Caveats}
To remain consistent with \citetalias{Ronco2024}, we computed the X-ray photoevaporation rates following \citet{Kunitomo2021}, who adopted a simplified version of the  \citet{Owen2012} prescription. As discussed in \citetalias{Ronco2024}, although \citet{Owen2012} rates might overestimate X-ray photoevaporation, they remain the only detailed prescriptions derived directly from hydrodynamic models that can be applied in our scenarios. While the full \citet{Owen2012} formulation yields shorter disk lifetimes and shifts evolutionary transitions to higher stellar masses, the overall qualitative behavior of disk dispersal remains robust. Crucially, more efficient photoevaporation rates deplete massive disks faster, which could hinder giant planet formation. In Appendix \ref{Ap:PhotoRates}, we test the \citet{Owen2012} rates exclusively on our nominal setup ($10\%$ disk mass, $\alpha_0=10^{-3}$, $t_{\text{start}}=0$~Myr, across all metallicities). These rates further shorten disk lifetimes around massive stars and cause a clear decline in giant production starting at $2 M_{\odot}$, shifting the occurrence rate peak to $\sim 2 M_{\odot}$, in better agreement with observations \citep{Jones2016, Wolthoff2022}.

We currently neglect external photoevaporation from nearby massive stars \citep{WinterHaworth2022}. Because this mechanism efficiently strips outer disk material, it would likely reduce giant planet formation around stars with $M_\star > 2\mathrm{M}_\odot$, which in our current models retain gas long enough (late gap-opening $>1$~Myr; \citetalias{Ronco2024}) to form giants efficiently. Including external photoevaporation would accelerate outside-in dispersal, prematurely halting the pebble flux toward the planet-forming zone, and naturally shift the theoretical occurrence peak closer to $\sim 2 \mathrm{M}_\odot$.

Regarding planet formation, we model a single planet per disk, neglecting the gravitational interactions inherent to multi-planet systems. Such N-body interactions can excite planetary eccentricities, modifying both migration pathways and pebble accretion rates, which would ultimately alter the final masses and orbital architectures of the gas giants.

Finally, rigorous statistical comparisons typically require a full population synthesis approach with initial parameters drawn from observed distributions \citep[e.g.,][]{johnston_2026}. We opted instead for a systematic physical exploration of gas and dust evolution coupled with migration. Remarkably, even without tuning our initial conditions to empirical distributions, our physical model naturally reproduces key observables: the rise in occurrence rates between $1$ and $2\,\mathrm{M}_\odot$, the subsequent sharp decline at higher masses, and the positive correlation with stellar metallicity. This detailed framework thus provides a robust physical foundation for future population synthesis studies.

\section{Conclusions}

The main goal of this work was to investigate the exoplanet populations that naturally emerge around intermediate-mass stars. We employed a comprehensive physical model that couples gas disk evolution, driven by pre-main-sequence stellar evolution, with dust growth and evolution, pebble and gas accretion, and detailed planetary migration. 

Through a systematic parameter study, we found that the occurrence rate of giant planets increases steadily with stellar mass from $1$ to $3 \mathrm{M}_\odot$, before dropping abruptly for stars with $M_\star \ge 4 \mathrm{M}_\odot$. This steep decline is a direct consequence of early stellar evolution, as intense FUV photoevaporation around massive stars drastically shortens disk lifetimes and halts planetary growth. Although our theoretical occurrence peak near $3 \mathrm{M}_\odot$ differs from the $\sim 2 \mathrm{M}_\odot$ peak inferred from radial velocity surveys \citep{Wolthoff2022}, our models successfully reproduce the observed positive correlation between giant planet occurrence and stellar metallicity. The mismatch in the peak mass may stem from the adopted initial conditions or the use of X-ray photoevaporation prescriptions that overestimate disk lifetimes. As discussed, adopting more efficient mass-loss rates, such as those by \citet{Owen2012}, could help reconcile our theoretical peak with current RV observations. Additionally, our simulated gas giants exhibit a period distribution shifted toward wider orbits. Driven by thermal torques, a large fraction of these planets settle beyond the detection limits of standard RV surveys. However, these wide-orbit companions represent prime targets for direct imaging, and their formation efficiency aligns well with the high occurrence rates reported by such surveys. These companions at large orbital separations should also survive the evolution of their host stars into white dwarfs and might be detectable at this stage with facilities such as JWST. 

Ultimately, our detailed physical framework naturally reproduces several key observables without requiring a full population synthesis approach. Using these robust formation models as the foundation for comprehensive population synthesis studies will be the focus of our future work.

\begin{acknowledgements}
MPR and MRS thank for support from ESO/comit{\'e} mixto who funded an extended research visit. 
MRS also acknowledges support from the research unit FOR2990 (eRO-STEP: Stellar endpoints with eROSITA) funded by the Deutsche Forschungsgemeinschaft and the internal research project PI\_LIR\_26\_07 at USM. MPR and OMG are partially supported by PIP 1122025010 0965CO from CONICET (Argentina). EV acknowledge support from the Spanish Ministry of Science and Innovation/State Agency of Research (MCIN/AEI) under the grant PID2021-127289-NB-I00. We also thank Juan Ignacio Rodriguez from IALP for the computation managing resources of the Grupo de Astrofísica Planetaria de La Plata. 

\end{acknowledgements}

\bibliographystyle{aa}
\bibliography{Biblio-PII}

@ARTICLE{emsenhuberetal21-1,
       author = {{Emsenhuber}, Alexandre and {Mordasini}, Christoph and {Burn}, Remo and {Alibert}, Yann and {Benz}, Willy and {Asphaug}, Erik},
        title = "{The New Generation Planetary Population Synthesis (NGPPS). I. Bern global model of planet formation and evolution, model tests, and emerging planetary systems}",
      journal = {\aap},
         year = 2021,
        month = dec,
       volume = {656},
          eid = {A69},
        pages = {A69},
          doi = {10.1051/0004-6361/202038553},
archivePrefix = {arXiv},
       eprint = {2007.05561},
 primaryClass = {astro-ph.EP},
       adsurl = {https://ui.adsabs.harvard.edu/abs/2021A&A...656A..69E}
}

@ARTICLE{Vigan_2021,
       author = {{Vigan}, A. and {Fontanive}, C. and {Meyer}, M. and {Biller}, B. and {Bonavita}, M. and {Feldt}, M. and {Desidera}, S. and {Marleau}, G.-D. and {Emsenhuber}, A. and {Galicher}, R. and {Rice}, K. and {Forgan}, D. and {Mordasini}, C. and {Gratton}, R. and {Le Coroller}, H. and {Maire}, A.-L. and {Cantalloube}, F. and {Chauvin}, G. and {Cheetham}, A. and {Hagelberg}, J. and {Lagrange}, A.-M. and {Langlois}, M. and {Bonnefoy}, M. and {Beuzit}, J.-L. and {Boccaletti}, A. and {D'Orazi}, V. and {Delorme}, P. and {Dominik}, C. and {Henning}, Th. and {Janson}, M. and {Lagadec}, E. and {Lazzoni}, C. and {Ligi}, R. and {Menard}, F. and {Mesa}, D. and {Messina}, S. and {Moutou}, C. and {M{\"u}ller}, A. and {Perrot}, C. and {Samland}, M. and {Schmid}, H.~M. and {Schmidt}, T. and {Sissa}, E. and {Turatto}, M. and {Udry}, S. and {Zurlo}, A. and {Abe}, L. and {Antichi}, J. and {Asensio-Torres}, R. and {Baruffolo}, A. and {Baudoz}, P. and {Baudrand}, J. and {Bazzon}, A. and {Blanchard}, P. and {Bohn}, A.~J. and {Brown Sevilla}, S. and {Carbillet}, M. and {Carle}, M. and {Cascone}, E. and {Charton}, J. and {Claudi}, R. and {Costille}, A. and {De Caprio}, V. and {Delboulb{\'e}}, A. and {Dohlen}, K. and {Engler}, N. and {Fantinel}, D. and {Feautrier}, P. and {Fusco}, T. and {Gigan}, P. and {Girard}, J.~H. and {Giro}, E. and {Gisler}, D. and {Gluck}, L. and {Gry}, C. and {Hubin}, N. and {Hugot}, E. and {Jaquet}, M. and {Kasper}, M. and {Le Mignant}, D. and {Llored}, M. and {Madec}, F. and {Magnard}, Y. and {Martinez}, P. and {Maurel}, D. and {M{\"o}ller-Nilsson}, O. and {Mouillet}, D. and {Moulin}, T. and {Orign{\'e}}, A. and {Pavlov}, A. and {Perret}, D. and {Petit}, C. and {Pragt}, J. and {Puget}, P. and {Rabou}, P. and {Ramos}, J. and {Rickman}, E.~L. and {Rigal}, F. and {Rochat}, S. and {Roelfsema}, R. and {Rousset}, G. and {Roux}, A. and {Salasnich}, B. and {Sauvage}, J.-F. and {Sevin}, A. and {Soenke}, C. and {Stadler}, E. and {Suarez}, M. and {Wahhaj}, Z. and {Weber}, L. and {Wildi}, F.},
        title = "{The SPHERE infrared survey for exoplanets (SHINE). III. The demographics of young giant exoplanets below 300 au with SPHERE}",
      journal = {\aap},
         year = 2021,
        month = jul,
       volume = {651},
          eid = {A72},
        pages = {A72},
          doi = {10.1051/0004-6361/202038107},
archivePrefix = {arXiv},
       eprint = {2007.06573},
 primaryClass = {astro-ph.EP},
       adsurl = {https://ui.adsabs.harvard.edu/abs/2021A&A...651A..72V}
}

@ARTICLE{Nielsen_2019,
       author = {{Nielsen}, Eric L. and {De Rosa}, Robert J. and {Macintosh}, Bruce and {Wang}, Jason J. and {Ruffio}, Jean-Baptiste and {Chiang}, Eugene and {Marley}, Mark S. and {Saumon}, Didier and {Savransky}, Dmitry and {Ammons}, S. Mark and {Bailey}, Vanessa P. and {Barman}, Travis and {Blain}, C{\'e}lia and {Bulger}, Joanna and {Burrows}, Adam and {Chilcote}, Jeffrey and {Cotten}, Tara and {Czekala}, Ian and {Doyon}, Rene and {Duch{\^e}ne}, Gaspard and {Esposito}, Thomas M. and {Fabrycky}, Daniel and {Fitzgerald}, Michael P. and {Follette}, Katherine B. and {Fortney}, Jonathan J. and {Gerard}, Benjamin L. and {Goodsell}, Stephen J. and {Graham}, James R. and {Greenbaum}, Alexandra Z. and {Hibon}, Pascale and {Hinkley}, Sasha and {Hirsch}, Lea A. and {Hom}, Justin and {Hung}, Li-Wei and {Dawson}, Rebekah Ilene and {Ingraham}, Patrick and {Kalas}, Paul and {Konopacky}, Quinn and {Larkin}, James E. and {Lee}, Eve J. and {Lin}, Jonathan W. and {Maire}, J{\'e}r{\^o}me and {Marchis}, Franck and {Marois}, Christian and {Metchev}, Stanimir and {Millar-Blanchaer}, Maxwell A. and {Morzinski}, Katie M. and {Oppenheimer}, Rebecca and {Palmer}, David and {Patience}, Jennifer and {Perrin}, Marshall and {Poyneer}, Lisa and {Pueyo}, Laurent and {Rafikov}, Roman R. and {Rajan}, Abhijith and {Rameau}, Julien and {Rantakyr{\"o}}, Fredrik T. and {Ren}, Bin and {Schneider}, Adam C. and {Sivaramakrishnan}, Anand and {Song}, Inseok and {Soummer}, Remi and {Tallis}, Melisa and {Thomas}, Sandrine and {Ward-Duong}, Kimberly and {Wolff}, Schuyler},
        title = "{The Gemini Planet Imager Exoplanet Survey: Giant Planet and Brown Dwarf Demographics from 10 to 100 au}",
      journal = {\aj},
         year = 2019,
        month = jul,
       volume = {158},
       number = {1},
          eid = {13},
        pages = {13},
          doi = {10.3847/1538-3881/ab16e9},
archivePrefix = {arXiv},
       eprint = {1904.05358},
 primaryClass = {astro-ph.EP},
       adsurl = {https://ui.adsabs.harvard.edu/abs/2019AJ....158...13N}
}

@ARTICLE{Currie_2022,
       author = {{Currie}, Thayne and {Lawson}, Kellen and {Schneider}, Glenn and {Lyra}, Wladimir and {Wisniewski}, John and {Grady}, Carol and {Guyon}, Olivier and {Tamura}, Motohide and {Kotani}, Takayuki and {Kawahara}, Hajime and {Brandt}, Timothy and {Uyama}, Taichi and {Muto}, Takayuki and {Dong}, Ruobing and {Kudo}, Tomoyuki and {Hashimoto}, Jun and {Fukagawa}, Misato and {Wagner}, Kevin and {Lozi}, Julien and {Chilcote}, Jeffrey and {Tobin}, Taylor and {Groff}, Tyler and {Ward-Duong}, Kimberly and {Januszewski}, William and {Norris}, Barnaby and {Tuthill}, Peter and {van der Marel}, Nienke and {Sitko}, Michael and {Deo}, Vincent and {Vievard}, Sebastien and {Jovanovic}, Nemanja and {Martinache}, Frantz and {Skaf}, Nour},
        title = "{Images of embedded Jovian planet formation at a wide separation around AB Aurigae}",
      journal = {Nature Astronomy},
         year = 2022,
        month = apr,
       volume = {6},
        pages = {751-759},
          doi = {10.1038/s41550-022-01634-x},
archivePrefix = {arXiv},
       eprint = {2204.00633},
 primaryClass = {astro-ph.EP},
       adsurl = {https://ui.adsabs.harvard.edu/abs/2022NatAs...6..751C}
}

@ARTICLE{Pinte_2018,
       author = {{Pinte}, C. and {Price}, D.~J. and {M{\'e}nard}, F. and {Duch{\^e}ne}, G. and {Dent}, W.~R.~F. and {Hill}, T. and {de Gregorio-Monsalvo}, I. and {Hales}, A. and {Mentiplay}, D.},
        title = "{Kinematic Evidence for an Embedded Protoplanet in a Circumstellar Disk}",
      journal = {\apjl},
         year = 2018,
        month = jun,
       volume = {860},
       number = {1},
          eid = {L13},
        pages = {L13},
          doi = {10.3847/2041-8213/aac6dc},
archivePrefix = {arXiv},
       eprint = {1805.10293},
 primaryClass = {astro-ph.SR},
       adsurl = {https://ui.adsabs.harvard.edu/abs/2018ApJ...860L..13P}
}

@ARTICLE{Quanz_2013,
       author = {{Quanz}, Sascha P. and {Amara}, Adam and {Meyer}, Michael R. and {Kenworthy}, Matthew A. and {Kasper}, Markus and {Girard}, Julien H.},
        title = "{A Young Protoplanet Candidate Embedded in the Circumstellar Disk of HD 100546}",
      journal = {\apjl},
         year = 2013,
        month = mar,
       volume = {766},
       number = {1},
          eid = {L1},
        pages = {L1},
          doi = {10.1088/2041-8205/766/1/L1},
archivePrefix = {arXiv},
       eprint = {1302.7122},
 primaryClass = {astro-ph.GA},
       adsurl = {https://ui.adsabs.harvard.edu/abs/2013ApJ...766L...1Q}
}

@ARTICLE{Andrews_2020,
       author = {{Andrews}, Sean M.},
        title = "{Observations of Protoplanetary Disk Structures}",
      journal = {\araa},
         year = 2020,
        month = aug,
       volume = {58},
        pages = {483-528},
          doi = {10.1146/annurev-astro-031220-010302},
archivePrefix = {arXiv},
       eprint = {2001.05007},
 primaryClass = {astro-ph.EP},
       adsurl = {https://ui.adsabs.harvard.edu/abs/2020ARA&A..58..483A}
}

@ARTICLE{Grant_2022,
       author = {{Grant}, Sierra L. and {Espaillat}, Catherine C. and {Brittain}, Sean and {Scott-Joseph}, Caleb and {Calvet}, Nuria},
        title = "{Tracing Accretion onto Herbig Ae/Be Stars Using the Br{\ensuremath{\gamma}} Line}",
      journal = {\apj},
         year = 2022,
        month = feb,
       volume = {926},
       number = {2},
          eid = {229},
        pages = {229},
          doi = {10.3847/1538-4357/ac450a},
archivePrefix = {arXiv},
       eprint = {2112.10428},
 primaryClass = {astro-ph.SR},
       adsurl = {https://ui.adsabs.harvard.edu/abs/2022ApJ...926..229G}
}

@ARTICLE{Wichittanakom_2020,
       author = {{Wichittanakom}, C. and {Oudmaijer}, R.~D. and {Fairlamb}, J.~R. and {Mendigut{\'\i}a}, I. and {Vioque}, M. and {Ababakr}, K.~M.},
        title = "{The accretion rates and mechanisms of Herbig Ae/Be stars}",
      journal = {\mnras},
         year = 2020,
        month = mar,
       volume = {493},
       number = {1},
        pages = {234-249},
          doi = {10.1093/mnras/staa169},
archivePrefix = {arXiv},
       eprint = {2001.05971},
 primaryClass = {astro-ph.SR},
       adsurl = {https://ui.adsabs.harvard.edu/abs/2020MNRAS.493..234W}
}

@ARTICLE{Fairlamb_2015,
       author = {{Fairlamb}, J.~R. and {Oudmaijer}, R.~D. and {Mendigut{\'\i}a}, I. and {Ilee}, J.~D. and {van den Ancker}, M.~E.},
        title = "{A spectroscopic survey of Herbig Ae/Be stars with X-shooter - I. Stellar parameters and accretion rates}",
      journal = {\mnras},
         year = 2015,
        month = oct,
       volume = {453},
       number = {1},
        pages = {976-1001},
          doi = {10.1093/mnras/stv1576},
archivePrefix = {arXiv},
       eprint = {1507.05967},
 primaryClass = {astro-ph.SR},
       adsurl = {https://ui.adsabs.harvard.edu/abs/2015MNRAS.453..976F}
}

@ARTICLE{Brittain_2023,
       author = {{Brittain}, Sean D. and {Kamp}, Inga and {Meeus}, Gwendolyn and {Oudmaijer}, Ren{\'e} D. and {Waters}, L.~B.~F.~M.},
        title = "{Herbig Stars}",
      journal = {\ssr},
         year = 2023,
        month = feb,
       volume = {219},
       number = {1},
          eid = {7},
        pages = {7},
          doi = {10.1007/s11214-023-00949-z},
archivePrefix = {arXiv},
       eprint = {2301.01165},
 primaryClass = {astro-ph.SR},
       adsurl = {https://ui.adsabs.harvard.edu/abs/2023SSRv..219....7B}
}

@ARTICLE{Brittain_2026,
       author = {{Brittain}, Sean D. and {Kern}, Joshua W. and {Meeus}, Gwendolyn and {Oudmaijer}, Ren{\'e} D.},
        title = "{Evolution of the Accretion Rate of Young Intermediate-mass Stars: Implications for Disk Evolution and Planet Formation}",
      journal = {\aj},
         year = 2026,
        month = jan,
       volume = {171},
       number = {1},
          eid = {4},
        pages = {4},
          doi = {10.3847/1538-3881/ae1a42},
archivePrefix = {arXiv},
       eprint = {2512.02237},
 primaryClass = {astro-ph.SR},
       adsurl = {https://ui.adsabs.harvard.edu/abs/2026AJ....171....4B}
}

@ARTICLE{Woelfer_2025,
       author = {{W{\"o}lfer}, Lisa and {Izquierdo}, Andr{\'e}s F. and {Booth}, Alice and {Facchini}, Stefano and {Teague}, Richard and {van Dishoeck}, Ewine F. and {Paneque-Carre{\~n}o}, Teresa and {Dent}, Bill},
        title = "{Spirals and Vertical Motions in the Planet-Forming Disk around HD 100546. A multi-line study of its gas kinematics}",
      journal = {arXiv e-prints},
         year = 2025,
        month = dec,
          eid = {arXiv:2512.13814},
        pages = {arXiv:2512.13814},
          doi = {10.48550/arXiv.2512.13814},
archivePrefix = {arXiv},
       eprint = {2512.13814},
 primaryClass = {astro-ph.EP},
       adsurl = {https://ui.adsabs.harvard.edu/abs/2025arXiv251213814W}
}

@ARTICLE{johnston2024,
       author = {{Johnston}, Heather F. and {Pani{\'c}}, O. and {Liu}, B.},
        title = "{Formation of giant planets around intermediate-mass stars}",
      journal = {\mnras},
         year = 2024,
        month = jan,
       volume = {527},
       number = {2},
        pages = {2303-2322},
          doi = {10.1093/mnras/stad3254},
archivePrefix = {arXiv},
       eprint = {2310.17767},
 primaryClass = {astro-ph.EP},
       adsurl = {https://ui.adsabs.harvard.edu/abs/2024MNRAS.527.2303J}
}

@ARTICLE{johnston_2025,
       author = {{Johnston}, Heather F. and {Pani{\'c}}, Olja and {Liu}, Beibei and {Jankowski}, Patryk},
        title = "{The impact of pre-main sequence stellar luminosity on giant planet formation}",
      journal = {\mnras},
         year = 2025,
        month = nov,
       volume = {543},
       number = {4},
        pages = {4053-4066},
          doi = {10.1093/mnras/staf1507},
archivePrefix = {arXiv},
       eprint = {2508.20291},
 primaryClass = {astro-ph.EP},
       adsurl = {https://ui.adsabs.harvard.edu/abs/2025MNRAS.543.4053J}
}

@ARTICLE{johnston_2026,
       author = {{Johnston}, Heather F and {Panic}, Olja and {Reffert}, Sabine and {Liu}, Beibei and {Ma}, Xinghao},
        title = "{Reproducing the stellar-mass dependence of the giant planet occurrence rate with pebble accretion models}",
      journal = {arXiv e-prints},
         year = 2026,
        month = mar,
          eid = {arXiv:2603.03014},
        pages = {arXiv:2603.03014},
          doi = {10.48550/arXiv.2603.03014},
archivePrefix = {arXiv},
       eprint = {2603.03014},
 primaryClass = {astro-ph.EP},
       adsurl = {https://ui.adsabs.harvard.edu/abs/2026arXiv260303014J}
}

@ARTICLE{fischer2005,
       author = {{Fischer}, Debra A. and {Valenti}, Jeff},
        title = "{The Planet-Metallicity Correlation}",
      journal = {\apj},
         year = 2005,
        month = apr,
       volume = {622},
       number = {2},
        pages = {1102-1117},
          doi = {10.1086/428383},
       adsurl = {https://ui.adsabs.harvard.edu/abs/2005ApJ...622.1102F}
}

@ARTICLE{Wolthoff2022,
       author = {{Wolthoff}, Vera and {Reffert}, Sabine and {Quirrenbach}, Andreas and {Jones}, Mat{\'\i}as I. and {Wittenmyer}, Robert A. and {Jenkins}, James S.},
        title = "{Precise radial velocities of giant stars. XVI. Planet occurrence rates from the combined analysis of the Lick, EXPRESS, and PPPS giant star surveys}",
      journal = {\aap},
         year = 2022,
        month = may,
       volume = {661},
          eid = {A63},
        pages = {A63},
          doi = {10.1051/0004-6361/202142501},
archivePrefix = {arXiv},
       eprint = {2202.12800},
 primaryClass = {astro-ph.EP},
       adsurl = {https://ui.adsabs.harvard.edu/abs/2022A&A...661A..63W}
}

@ARTICLE{Gorti2009,
       author = {{Gorti}, U. and {Dullemond}, C.~P. and {Hollenbach}, D.},
        title = "{Time Evolution of Viscous Circumstellar Disks due to Photoevaporation by Far-Ultraviolet, Extreme-Ultraviolet, and X-ray Radiation from the Central Star}",
      journal = {\apj},
         year = 2009,
        month = nov,
       volume = {705},
       number = {2},
        pages = {1237-1251},
          doi = {10.1088/0004-637X/705/2/1237},
archivePrefix = {arXiv},
       eprint = {0909.1836},
 primaryClass = {astro-ph.SR},
       adsurl = {https://ui.adsabs.harvard.edu/abs/2009ApJ...705.1237G}
}

@ARTICLE{Jura2003,
       author = {{Jura}, M.},
        title = "{A Tidally Disrupted Asteroid around the White Dwarf G29-38}",
      journal = {\apjl},
         year = 2003,
        month = feb,
       volume = {584},
       number = {2},
        pages = {L91-L94},
          doi = {10.1086/374036},
archivePrefix = {arXiv},
       eprint = {astro-ph/0301411},
 primaryClass = {astro-ph},
       adsurl = {https://ui.adsabs.harvard.edu/abs/2003ApJ...584L..91J}
}

@ARTICLE{Gaensicke2006,
       author = {{G{\"a}nsicke}, B.~T. and {Marsh}, T.~R. and {Southworth}, J. and {Rebassa-Mansergas}, A.},
        title = "{A Gaseous Metal Disk Around a White Dwarf}",
      journal = {Science},
         year = 2006,
        month = dec,
       volume = {314},
       number = {5807},
        pages = {1908},
          doi = {10.1126/science.1135033},
archivePrefix = {arXiv},
       eprint = {astro-ph/0612697},
 primaryClass = {astro-ph},
       adsurl = {https://ui.adsabs.harvard.edu/abs/2006Sci...314.1908G}
}

@ARTICLE{Gaensicke2019,
       author = {{G{\"a}nsicke}, Boris T. and {Schreiber}, Matthias R. and {Toloza}, Odette and {Gentile Fusillo}, Nicola P. and {Koester}, Detlev and {Manser}, Christopher J.},
        title = "{Accretion of a giant planet onto a white dwarf star}",
      journal = {\nat},
         year = 2019,
        month = dec,
       volume = {576},
       number = {7785},
        pages = {61-64},
          doi = {10.1038/s41586-019-1789-8},
archivePrefix = {arXiv},
       eprint = {1912.01611},
 primaryClass = {astro-ph.EP},
       adsurl = {https://ui.adsabs.harvard.edu/abs/2019Natur.576...61G}
}

@ARTICLE{Vandenburg2020,
       author = {{Vanderburg}, Andrew and {Rappaport}, Saul A. and {Xu}, Siyi and {Crossfield}, Ian J.~M. and {Becker}, Juliette C. and {Gary}, Bruce and {Murgas}, Felipe and {Blouin}, Simon and {Kaye}, Thomas G. and {Palle}, Enric and {Melis}, Carl and {Morris}, Brett M. and {Kreidberg}, Laura and {Gorjian}, Varoujan and {Morley}, Caroline V. and {Mann}, Andrew W. and {Parviainen}, Hannu and {Pearce}, Logan A. and {Newton}, Elisabeth R. and {Carrillo}, Andreia and {Zuckerman}, Ben and {Nelson}, Lorne and {Zeimann}, Greg and {Brown}, Warren R. and {Tronsgaard}, Ren{\'e} and {Klein}, Beth and {Ricker}, George R. and {Vanderspek}, Roland K. and {Latham}, David W. and {Seager}, Sara and {Winn}, Joshua N. and {Jenkins}, Jon M. and {Adams}, Fred C. and {Benneke}, Bj{\"o}rn and {Berardo}, David and {Buchhave}, Lars A. and {Caldwell}, Douglas A. and {Christiansen}, Jessie L. and {Collins}, Karen A. and {Col{\'o}n}, Knicole D. and {Daylan}, Tansu and {Doty}, John and {Doyle}, Alexandra E. and {Dragomir}, Diana and {Dressing}, Courtney and {Dufour}, Patrick and {Fukui}, Akihiko and {Glidden}, Ana and {Guerrero}, Natalia M. and {Guo}, Xueying and {Heng}, Kevin and {Henriksen}, Andreea I. and {Huang}, Chelsea X. and {Kaltenegger}, Lisa and {Kane}, Stephen R. and {Lewis}, John A. and {Lissauer}, Jack J. and {Morales}, Farisa and {Narita}, Norio and {Pepper}, Joshua and {Rose}, Mark E. and {Smith}, Jeffrey C. and {Stassun}, Keivan G. and {Yu}, Liang},
        title = "{A giant planet candidate transiting a white dwarf}",
      journal = {\nat},
         year = 2020,
        month = sep,
       volume = {585},
       number = {7825},
        pages = {363-367},
          doi = {10.1038/s41586-020-2713-y},
archivePrefix = {arXiv},
       eprint = {2009.07282},
 primaryClass = {astro-ph.EP},
       adsurl = {https://ui.adsabs.harvard.edu/abs/2020Natur.585..363V}
}

@ARTICLE{Koester2014,
       author = {{Koester}, D. and {G{\"a}nsicke}, B.~T. and {Farihi}, J.},
        title = "{The frequency of planetary debris around young white dwarfs}",
      journal = {\aap},
         year = 2014,
        month = jun,
       volume = {566},
          eid = {A34},
        pages = {A34},
          doi = {10.1051/0004-6361/201423691},
archivePrefix = {arXiv},
       eprint = {1404.2617},
 primaryClass = {astro-ph.SR},
       adsurl = {https://ui.adsabs.harvard.edu/abs/2014A&A...566A..34K}
}

@ARTICLE{Blackman2021,
       author = {{Blackman}, J.~W. and {Beaulieu}, J.~P. and {Bennett}, D.~P. and {Danielski}, C. and {Alard}, C. and {Cole}, A.~A. and {Vandorou}, A. and {Ranc}, C. and {Terry}, S.~K. and {Bhattacharya}, A. and {Bond}, I. and {Bachelet}, E. and {Veras}, D. and {Koshimoto}, N. and {Batista}, V. and {Marquette}, J.~B.},
        title = "{A Jovian analogue orbiting a white dwarf star}",
      journal = {\nat},
         year = 2021,
        month = oct,
       volume = {598},
       number = {7880},
        pages = {272-275},
          doi = {10.1038/s41586-021-03869-6},
archivePrefix = {arXiv},
       eprint = {2110.07934},
 primaryClass = {astro-ph.EP},
       adsurl = {https://ui.adsabs.harvard.edu/abs/2021Natur.598..272B}
}

@ARTICLE{Ronco2020,
       author = {{Ronco}, Mar{\'\i}a Paula and {Schreiber}, Matthias R. and {Giuppone}, Cristian A. and {Veras}, Dimitri and {Cuadra}, Jorge and {Guilera}, Octavio M.},
        title = "{How Jupiters Save or Destroy Inner Neptunes around Evolved Stars}",
      journal = {\apjl},
         year = 2020,
        month = jul,
       volume = {898},
       number = {1},
          eid = {L23},
        pages = {L23},
          doi = {10.3847/2041-8213/aba35f},
archivePrefix = {arXiv},
       eprint = {2007.04337},
 primaryClass = {astro-ph.EP},
       adsurl = {https://ui.adsabs.harvard.edu/abs/2020ApJ...898L..23R}
}

@ARTICLE{Manser2019,
       author = {{Manser}, Christopher J. and {G{\"a}nsicke}, Boris T. and {Eggl}, Siegfried and {Hollands}, Mark and {Izquierdo}, Paula and {Koester}, Detlev and {Landstreet}, John D. and {Lyra}, Wladimir and {Marsh}, Thomas R. and {Meru}, Farzana and {Mustill}, Alexander J. and {Rodr{\'\i}guez-Gil}, Pablo and {Toloza}, Odette and {Veras}, Dimitri and {Wilson}, David J. and {Burleigh}, Matthew R. and {Davies}, Melvyn B. and {Farihi}, Jay and {Gentile Fusillo}, Nicola and {de Martino}, Domitilla and {Parsons}, Steven G. and {Quirrenbach}, Andreas and {Raddi}, Roberto and {Reffert}, Sabine and {Del Santo}, Melania and {Schreiber}, Matthias R. and {Silvotti}, Roberto and {Toonen}, Silvia and {Villaver}, Eva and {Wyatt}, Mark and {Xu}, Siyi and {Portegies Zwart}, Simon},
        title = "{A planetesimal orbiting within the debris disc around a white dwarf star}",
      journal = {Science},
         year = 2019,
        month = apr,
       volume = {364},
       number = {6435},
        pages = {66-69},
          doi = {10.1126/science.aat5330},
archivePrefix = {arXiv},
       eprint = {1904.02163},
 primaryClass = {astro-ph.EP},
       adsurl = {https://ui.adsabs.harvard.edu/abs/2019Sci...364...66M}
}

@ARTICLE{Venturini2020SE,
       author = {{Venturini}, Julia and {Guilera}, Octavio Miguel and {Ronco}, Mar{\'\i}a Paula and {Mordasini}, Christoph},
        title = "{Most super-Earths formed by dry pebble accretion are less massive than 5 Earth masses}",
      journal = {\aap},
         year = 2020,
        month = dec,
       volume = {644},
          eid = {A174},
        pages = {A174},
          doi = {10.1051/0004-6361/202039140},
archivePrefix = {arXiv},
       eprint = {2008.05497},
 primaryClass = {astro-ph.EP},
       adsurl = {https://ui.adsabs.harvard.edu/abs/2020A&A...644A.174V}
}

@ARTICLE{Venturini2020Letter,
       author = {{Venturini}, Julia and {Guilera}, Octavio M. and {Haldemann}, Jonas and {Ronco}, Mar{\'\i}a P. and {Mordasini}, Christoph},
        title = "{The nature of the radius valley. Hints from formation and evolution models}",
      journal = {\aap},
         year = 2020,
        month = nov,
       volume = {643},
          eid = {L1},
        pages = {L1},
          doi = {10.1051/0004-6361/202039141},
archivePrefix = {arXiv},
       eprint = {2008.05513},
 primaryClass = {astro-ph.EP},
       adsurl = {https://ui.adsabs.harvard.edu/abs/2020A&A...643L...1V}
}

@article{Birnstiel11,
	Adsurl = {https://ui.adsabs.harvard.edu/abs/2011A&A...525A..11B},
	Archiveprefix = {arXiv},
	Author = {{Birnstiel}, T. and {Ormel}, C.~W. and {Dullemond}, C.~P.},
	Doi = {10.1051/0004-6361/201015228},
	Eid = {A11},
	Eprint = {1009.3011},
	Journal = {\aap},
	Month = jan,
	Pages = {A11},
	Primaryclass = {astro-ph.EP},
	Title = {{Dust size distributions in coagulation/fragmentation equilibrium: numerical solutions and analytical fits}},
	Volume = {525},
	Year = 2011}

@article{Drazkowska17,
	Adsurl = {https://ui.adsabs.harvard.edu/abs/2017A&A...608A..92D},
	Archiveprefix = {arXiv},
	Author = {{Dr{\k{a}}{\.z}kowska}, J. and {Alibert}, Y.},
	Doi = {10.1051/0004-6361/201731491},
	Eid = {A92},
	Eprint = {1710.00009},
	Journal = {\aap},
	Month = dec,
	Pages = {A92},
	Primaryclass = {astro-ph.EP},
	Title = {{Planetesimal formation starts at the snow line}},
	Volume = {608},
	Year = 2017}

@article{Tanigawa07,
	Adsurl = {http://adsabs.harvard.edu/abs/2007ApJ...667..557T},
	Archiveprefix = {arXiv},
	Author = {{Tanigawa}, T. and {Ikoma}, M.},
	Doi = {10.1086/520499},
	Eprint = {0705.4343},
	Journal = {\apj},
	Month = sep,
	Pages = {557-570},
	Title = {{A Systematic Study of the Final Masses of Gas Giant Planets}},
	Volume = 667,
	Year = 2007}

@article{Birnstiel12,
	Adsurl = {http://adsabs.harvard.edu/abs/2012A%26A...539A.148B},
	Archiveprefix = {arXiv},
	Author = {{Birnstiel}, T. and {Klahr}, H. and {Ercolano}, B.},
	Doi = {10.1051/0004-6361/201118136},
	Eid = {A148},
	Eprint = {1201.5781},
	Journal = {\aap},
	Month = mar,
	Pages = {A148},
	Primaryclass = {astro-ph.EP},
	Title = {{A simple model for the evolution of the dust population in protoplanetary disks}},
	Volume = 539,
	Year = 2012}

@article{Drazkowska16,
	Adsurl = {http://adsabs.harvard.edu/abs/2016A%26A...594A.105D},
	Archiveprefix = {arXiv},
	Author = {{Dr{\c a}zkowska}, J. and {Alibert}, Y. and {Moore}, B.},
	Doi = {10.1051/0004-6361/201628983},
	Eid = {A105},
	Eprint = {1607.05734},
	Journal = {\aap},
	Month = oct,
	Pages = {A105},
	Primaryclass = {astro-ph.EP},
	Title = {{Close-in planetesimal formation by pile-up of drifting pebbles}},
	Volume = 594,
	Year = 2016}

@article{Weiden77,
	Adsurl = {http://adsabs.harvard.edu/abs/1977MNRAS.180...57W},
	Author = {{Weidenschilling}, S.~J.},
	Journal = {\mnras},
	Month = jul,
	Pages = {57-70},
	Title = {{Aerodynamics of solid bodies in the solar nebula}},
	Volume = 180,
	Year = 1977}

@article{Tanaka02,
	Adsurl = {http://adsabs.harvard.edu/abs/2002ApJ...565.1257T},
	Author = {{Tanaka}, H. and {Takeuchi}, T. and {Ward}, W.~R.},
	Doi = {10.1086/324713},
	Journal = {\apj},
	Month = feb,
	Pages = {1257-1274},
	Title = {{Three-Dimensional Interaction between a Planet and an Isothermal Gaseous Disk. I. Corotation and Lindblad Torques and Planet Migration}},
	Volume = 565,
	Year = 2002}

@article{Crida2006,
	Adsurl = {http://adsabs.harvard.edu/abs/2006Icar..181..587C},
	Author = {{Crida}, A. and {Morbidelli}, A. and {Masset}, F.},
	Doi = {10.1016/j.icarus.2005.10.007},
	Eprint = {arXiv:astro-ph/0511082},
	Journal = {\icarus},
	Month = apr,
	Pages = {587-604},
	Title = {{On the width and shape of gaps in protoplanetary disks}},
	Volume = 181,
	Year = 2006}

@article{Mayor2011,
	Adsurl = {http://adsabs.harvard.edu/abs/2011arXiv1109.2497M},
	Archiveprefix = {arXiv},
	Author = {{Mayor}, M. and {Marmier}, M. and {Lovis}, C. and {Udry}, S. and {S{\'e}gransan}, D. and {Pepe}, F. and {Benz}, W. and {Bertaux}, J.~-. and {Bouchy}, F. and {Dumusque}, X. and {Lo Curto}, G. and {Mordasini}, C. and {Queloz}, D. and {Santos}, N.~C.},
	Eprint = {1109.2497},
	Journal = {ArXiv Astrophysics e-prints},
	Month = sep,
	Primaryclass = {astro-ph.EP},
	Title = {{The HARPS search for southern extra-solar planets XXXIV. Occurrence, mass distribution and orbital properties of super-Earths and Neptune-mass planets}},
	Year = 2011}

@article{Lodders09,
	Adsurl = {http://adsabs.harvard.edu/abs/2009LanB...4B...44L},
	Archiveprefix = {arXiv},
	Author = {{Lodders}, K. and {Palme}, H. and {Gail}, H.-P.},
	Doi = {10.1007/978-3-540-88055-4_34},
	Eprint = {0901.1149},
	Journal = {Landolt B{\"o}rnstein},
	Pages = {44},
	Primaryclass = {astro-ph.EP},
	Title = {{Abundances of the Elements in the Solar System}},
	Year = 2009}

@article{Ronco2017,
	Adsurl = {https://ui.adsabs.harvard.edu/abs/2017MNRAS.471.2753R},
	Archiveprefix = {arXiv},
	Author = {{Ronco}, M.~P. and {Guilera}, O.~M. and {de El{\'\i}a}, G.~C.},
	Doi = {10.1093/mnras/stx1746},
	Eprint = {1705.08608},
	Journal = {\mnras},
	Month = {Nov},
	Number = {3},
	Pages = {2753-2770},
	Primaryclass = {astro-ph.EP},
	Title = {{Formation of solar system analogues - I. Looking for initial conditions through a population synthesis analysis}},
	Volume = {471},
	Year = {2017}}

@article{Guilera2017,
	Adsurl = {https://ui.adsabs.harvard.edu/abs/2017A&A...604A..10G},
	Archiveprefix = {arXiv},
	Author = {{Guilera}, O.~M. and {S{\'a}ndor}, Zs.},
	Doi = {10.1051/0004-6361/201629843},
	Eid = {A10},
	Eprint = {1610.01232},
	Journal = {\aap},
	Month = {Jul},
	Pages = {A10},
	Primaryclass = {astro-ph.EP},
	Title = {{Giant planet formation at the pressure maxima of protoplanetary disks}},
	Volume = {604},
	Year = {2017}}

@article{JM17,
	Adsurl = {http://adsabs.harvard.edu/abs/2017MNRAS.471.4917J},
	Author = {{Jim{\'e}nez}, M.~A. and {Masset}, F.~S.},
	Doi = {10.1093/mnras/stx1946},
	Journal = {\mnras},
	Month = nov,
	Pages = {4917-4929},
	Title = {{Improved torque formula for low- and intermediate-mass planetary migration}},
	Volume = 471,
	Year = 2017}

@article{Guilera2019,
	Adsurl = {https://ui.adsabs.harvard.edu/abs/2019MNRAS.486.5690G},
	Archiveprefix = {arXiv},
	Author = {{Guilera}, O.~M. and {Cuello}, N. and {Montesinos}, M. and {Miller Bertolami}, M.~M. and {Ronco}, M.~P. and {Cuadra}, J. and {Masset}, F.~S.},
	Doi = {10.1093/mnras/stz1158},
	Eprint = {1904.11047},
	Journal = {\mnras},
	Month = {Jul},
	Number = {4},
	Pages = {5690-5708},
	Primaryclass = {astro-ph.EP},
	Title = {{Thermal torque effects on the migration of growing low-mass planets}},
	Volume = {486},
	Year = {2019}}

@article{Benitez-llambay2015,
	Adsurl = {http://adsabs.harvard.edu/abs/2015Natur.520...63B},
	Archiveprefix = {arXiv},
	Author = {{Ben{\'{\i}}tez-Llambay}, P. and {Masset}, F. and {Koenigsberger}, G. and {Szul{\'a}gyi}, J.},
	Doi = {10.1038/nature14277},
	Eprint = {1510.01778},
	Journal = {\nat},
	Month = apr,
	Pages = {63-65},
	Primaryclass = {astro-ph.EP},
	Title = {{Planet heating prevents inward migration of planetary cores}},
	Volume = 520,
	Year = 2015}

@article{masset2017,
	Adsurl = {http://adsabs.harvard.edu/abs/2017MNRAS.472.4204M},
	Author = {{Masset}, F.~S.},
	Doi = {10.1093/mnras/stx2271},
	Journal = {\mnras},
	Month = dec,
	Pages = {4204-4219},
	Title = {{Coorbital thermal torques on low-mass protoplanets}},
	Volume = 472,
	Year = 2017}

@article{Owen2012,
	Adsurl = {https://ui.adsabs.harvard.edu/abs/2012MNRAS.422.1880O},
	Archiveprefix = {arXiv},
	Author = {{Owen}, James E. and {Clarke}, Cathie J. and {Ercolano}, Barbara},
	Doi = {10.1111/j.1365-2966.2011.20337.x},
	Eprint = {1112.1087},
	Journal = {\mnras},
	Month = {May},
	Number = {3},
	Pages = {1880-1901},
	Primaryclass = {astro-ph.SR},
	Title = {{On the theory of disc photoevaporation}},
	Volume = {422},
	Year = {2012}}

@article{Brasser2017,
	Adsurl = {https://ui.adsabs.harvard.edu/abs/2017E&PSL.468...85B},
	Archiveprefix = {arXiv},
	Author = {{Brasser}, R. and {Mojzsis}, S.~J. and {Matsumura}, S. and {Ida}, S.},
	Doi = {10.1016/j.epsl.2017.04.005},
	Eprint = {1704.00184},
	Journal = {Earth and Planetary Science Letters},
	Month = {Jun},
	Pages = {85-93},
	Primaryclass = {astro-ph.EP},
	Title = {{The cool and distant formation of Mars}},
	Volume = {468},
	Year = {2017}}

@ARTICLE{Guilera2020,
       author = {{Guilera}, Octavio Miguel and {S{\'a}ndor}, Zsolt and {Ronco}, Mar{\'\i}a Paula and {Venturini}, Julia and {Miller Bertolami}, Marcelo Miguel},
        title = "{Giant planet formation at the pressure maxima of protoplanetary disks. II. A hybrid accretion scenario}",
      journal = {\aap},
         year = 2020,
        month = oct,
       volume = {642},
          eid = {A140},
        pages = {A140},
          doi = {10.1051/0004-6361/202038458},
archivePrefix = {arXiv},
       eprint = {2005.10868},
 primaryClass = {astro-ph.EP},
       adsurl = {https://ui.adsabs.harvard.edu/abs/2020A&A...642A.140G}
}

@ARTICLE{YoudinLithwick2007,
       author = {{Youdin}, Andrew N. and {Lithwick}, Yoram},
        title = "{Particle stirring in turbulent gas disks: Including orbital oscillations}",
      journal = {\icarus},
         year = 2007,
        month = dec,
       volume = {192},
       number = {2},
        pages = {588-604},
          doi = {10.1016/j.icarus.2007.07.012},
archivePrefix = {arXiv},
       eprint = {0707.2975},
 primaryClass = {astro-ph},
       adsurl = {https://ui.adsabs.harvard.edu/abs/2007Icar..192..588Y}
}

@ARTICLE{Kunitomo2021,
       author = {{Kunitomo}, Masanobu and {Ida}, Shigeru and {Takeuchi}, Taku and {Pani{\'c}}, Olja and {Miley}, James M. and {Suzuki}, Takeru K.},
        title = "{Photoevaporative Dispersal of Protoplanetary Disks around Evolving Intermediate-mass Stars}",
      journal = {\apj},
         year = 2021,
        month = mar,
       volume = {909},
       number = {2},
          eid = {109},
        pages = {109},
          doi = {10.3847/1538-4357/abdb2a10.48550/arXiv.2103.07673},
archivePrefix = {arXiv},
       eprint = {2103.07673},
 primaryClass = {astro-ph.EP},
       adsurl = {https://ui.adsabs.harvard.edu/abs/2021ApJ...909..109K}
}

@article{Andrews2010,
	Adsurl = {http://adsabs.harvard.edu/abs/2010ApJ...723.1241A},
	Archiveprefix = {arXiv},
	Author = {{Andrews}, S.~M. and {Wilner}, D.~J. and {Hughes}, A.~M. and {Qi}, C. and {Dullemond}, C.~P.},
	Doi = {10.1088/0004-637X/723/2/1241},
	Eprint = {1007.5070},
	Journal = {\apj},
	Month = nov,
	Pages = {1241-1254},
	Primaryclass = {astro-ph.SR},
	Title = {{Protoplanetary Disk Structures in Ophiuchus. II. Extension to Fainter Sources}},
	Volume = 723,
	Year = 2010}

@ARTICLE{Guilera2021,
       author = {{Guilera}, Octavio M. and {Miller Bertolami}, Marcelo M. and {Masset}, Frederic and {Cuadra}, Jorge and {Venturini}, Julia and {Ronco}, Mar{\'\i}a P.},
        title = "{The importance of thermal torques on the migration of planets growing by pebble accretion}",
      journal = {\mnras},
         year = 2021,
        month = nov,
       volume = {507},
       number = {3},
        pages = {3638-3652},
          doi = {10.1093/mnras/stab2371},
archivePrefix = {arXiv},
       eprint = {2108.04880},
 primaryClass = {astro-ph.EP},
       adsurl = {https://ui.adsabs.harvard.edu/abs/2021MNRAS.507.3638G}
}

@ARTICLE{Burn2021,
       author = {{Burn}, R. and {Schlecker}, M. and {Mordasini}, C. and {Emsenhuber}, A. and {Alibert}, Y. and {Henning}, T. and {Klahr}, H. and {Benz}, W.},
        title = "{The New Generation Planetary Population Synthesis (NGPPS). IV. Planetary systems around low-mass stars}",
      journal = {\aap},
         year = 2021,
        month = dec,
       volume = {656},
          eid = {A72},
        pages = {A72},
          doi = {10.1051/0004-6361/202140390},
archivePrefix = {arXiv},
       eprint = {2105.04596},
 primaryClass = {astro-ph.EP},
       adsurl = {https://ui.adsabs.harvard.edu/abs/2021A&A...656A..72B}
}

@ARTICLE{Alibert2011,
       author = {{Alibert}, Y. and {Mordasini}, C. and {Benz}, W.},
        title = "{Extrasolar planet population synthesis. III. Formation of planets around stars of different masses}",
      journal = {\aap},
         year = 2011,
        month = feb,
       volume = {526},
          eid = {A63},
        pages = {A63},
          doi = {10.1051/0004-6361/201014760},
archivePrefix = {arXiv},
       eprint = {1101.0513},
 primaryClass = {astro-ph.EP},
       adsurl = {https://ui.adsabs.harvard.edu/abs/2011A&A...526A..63A}
}

@ARTICLE{Cumming2008,
       author = {{Cumming}, Andrew and {Butler}, R. Paul and {Marcy}, Geoffrey W. and {Vogt}, Steven S. and {Wright}, Jason T. and {Fischer}, Debra A.},
        title = "{The Keck Planet Search: Detectability and the Minimum Mass and Orbital Period Distribution of Extrasolar Planets}",
      journal = {\pasp},
         year = 2008,
        month = may,
       volume = {120},
       number = {867},
        pages = {531},
          doi = {10.1086/588487},
archivePrefix = {arXiv},
       eprint = {0803.3357},
 primaryClass = {astro-ph},
       adsurl = {https://ui.adsabs.harvard.edu/abs/2008PASP..120..531C}
}

@ARTICLE{Borgniet2019,
       author = {{Borgniet}, S. and {Lagrange}, A. -M. and {Meunier}, N. and {Galland}, F. and {Arnold}, L. and {Astudillo-Defru}, N. and {Beuzit}, J. -L. and {Boisse}, I. and {Bonfils}, X. and {Bouchy}, F. and {Debondt}, K. and {Deleuil}, M. and {Delfosse}, X. and {Desort}, M. and {D{\'\i}az}, R.~F. and {Eggenberger}, A. and {Ehrenreich}, D. and {Forveille}, T. and {H{\'e}brard}, G. and {Loeillet}, B. and {Lovis}, C. and {Montagnier}, G. and {Moutou}, C. and {Pepe}, F. and {Perrier}, C. and {Pont}, F. and {Queloz}, D. and {Santerne}, A. and {Santos}, N.~C. and {S{\'e}gransan}, D. and {da Silva}, R. and {Sivan}, J.~P. and {Udry}, S. and {Vidal-Madjar}, A.},
        title = "{Extrasolar planets and brown dwarfs around AF-type stars. X. The SOPHIE sample: combining the SOPHIE and HARPS surveys to compute the close giant planet mass-period distribution around AF-type stars}",
      journal = {\aap},
         year = 2019,
        month = jan,
       volume = {621},
          eid = {A87},
        pages = {A87},
          doi = {10.1051/0004-6361/201833431},
archivePrefix = {arXiv},
       eprint = {1809.09914},
 primaryClass = {astro-ph.EP},
       adsurl = {https://ui.adsabs.harvard.edu/abs/2019A&A...621A..87B}
}

@ARTICLE{KennedyKenyon2008,
       author = {{Kennedy}, Grant M. and {Kenyon}, Scott J.},
        title = "{Planet Formation around Stars of Various Masses: The Snow Line and the Frequency of Giant Planets}",
      journal = {\apj},
         year = 2008,
        month = jan,
       volume = {673},
       number = {1},
        pages = {502-512},
          doi = {10.1086/524130},
archivePrefix = {arXiv},
       eprint = {0710.1065},
 primaryClass = {astro-ph},
       adsurl = {https://ui.adsabs.harvard.edu/abs/2008ApJ...673..502K}
}

@ARTICLE{MustillVillaver2012,
       author = {{Mustill}, Alexander J. and {Villaver}, Eva},
        title = "{Foretellings of Ragnar{\"o}k: World-engulfing Asymptotic Giants and the Inheritance of White Dwarfs}",
      journal = {\apj},
         year = 2012,
        month = dec,
       volume = {761},
       number = {2},
          eid = {121},
        pages = {121},
          doi = {10.1088/0004-637X/761/2/121},
archivePrefix = {arXiv},
       eprint = {1210.0328},
 primaryClass = {astro-ph.EP},
       adsurl = {https://ui.adsabs.harvard.edu/abs/2012ApJ...761..121M}
}

@ARTICLE{Reffert2015,
       author = {{Reffert}, Sabine and {Bergmann}, Christoph and {Quirrenbach}, Andreas and {Trifonov}, Trifon and {K{\"u}nstler}, Andreas},
        title = "{Precise radial velocities of giant stars. VII. Occurrence rate of giant extrasolar planets as a function of mass and metallicity}",
      journal = {\aap},
         year = 2015,
        month = feb,
       volume = {574},
          eid = {A116},
        pages = {A116},
          doi = {10.1051/0004-6361/201322360},
archivePrefix = {arXiv},
       eprint = {1412.4634},
 primaryClass = {astro-ph.EP},
       adsurl = {https://ui.adsabs.harvard.edu/abs/2015A&A...574A.116R}
}

@ARTICLE{Tremblay2016,
       author = {{Tremblay}, P. -E. and {Cummings}, J. and {Kalirai}, J.~S. and {G{\"a}nsicke}, B.~T. and {Gentile-Fusillo}, N. and {Raddi}, R.},
        title = "{The field white dwarf mass distribution}",
      journal = {\mnras},
         year = 2016,
        month = sep,
       volume = {461},
       number = {2},
        pages = {2100-2114},
          doi = {10.1093/mnras/stw1447},
archivePrefix = {arXiv},
       eprint = {1606.05292},
 primaryClass = {astro-ph.SR},
       adsurl = {https://ui.adsabs.harvard.edu/abs/2016MNRAS.461.2100T}
}

@ARTICLE{Cummings2018,
       author = {{Cummings}, Jeffrey D. and {Kalirai}, Jason S. and {Tremblay}, P. -E. and {Ramirez-Ruiz}, Enrico and {Choi}, Jieun},
        title = "{The White Dwarf Initial-Final Mass Relation for Progenitor Stars from 0.85 to 7.5 M $_{{\ensuremath{\odot}}}$}",
      journal = {\apj},
         year = 2018,
        month = oct,
       volume = {866},
       number = {1},
          eid = {21},
        pages = {21},
          doi = {10.3847/1538-4357/aadfd6},
archivePrefix = {arXiv},
       eprint = {1809.01673},
 primaryClass = {astro-ph.SR},
       adsurl = {https://ui.adsabs.harvard.edu/abs/2018ApJ...866...21C}
}

@ARTICLE{Johnson2010,
       author = {{Johnson}, John Asher and {Aller}, Kimberly M. and {Howard}, Andrew W. and {Crepp}, Justin R.},
        title = "{Giant Planet Occurrence in the Stellar Mass-Metallicity Plane}",
      journal = {\pasp},
         year = 2010,
        month = aug,
       volume = {122},
       number = {894},
        pages = {905},
          doi = {10.1086/655775},
archivePrefix = {arXiv},
       eprint = {1005.3084},
 primaryClass = {astro-ph.EP},
       adsurl = {https://ui.adsabs.harvard.edu/abs/2010PASP..122..905J}
}

@ARTICLE{Veras2014,
       author = {{Veras}, Dimitri and {Leinhardt}, Zo{\"e} M. and {Bonsor}, Amy and
         {G{\"a}nsicke}, Boris T.},
        title = "{Formation of planetary debris discs around white dwarfs - I. Tidal disruption of an extremely eccentric asteroid}",
      journal = {\mnras},
         year = 2014,
        month = dec,
       volume = {445},
       number = {3},
        pages = {2244-2255},
          doi = {10.1093/mnras/stu1871},
archivePrefix = {arXiv},
       eprint = {1409.2493},
 primaryClass = {astro-ph.EP},
       adsurl = {https://ui.adsabs.harvard.edu/abs/2014MNRAS.445.2244V}
}

@ARTICLE{Ikoma2000,
       author = {{Ikoma}, Masahiro and {Nakazawa}, Kiyoshi and {Emori}, Hiroyuki},
        title = "{Formation of Giant Planets: Dependences on Core Accretion Rate and Grain Opacity}",
      journal = {\apj},
         year = 2000,
        month = jul,
       volume = {537},
       number = {2},
        pages = {1013-1025},
          doi = {10.1086/309050},
       adsurl = {https://ui.adsabs.harvard.edu/abs/2000ApJ...537.1013I}
}

@INCOLLECTION{Veras2021,
       author = {{Veras}, Dimitri},
        title = "{Planetary Systems Around White Dwarfs}",
    booktitle = {Oxford Research Encyclopedia of Planetary Science},
         year = 2021,
          eid = {1},
        pages = {1},
          doi = {10.1093/acrefore/9780190647926.013.238},
       adsurl = {https://ui.adsabs.harvard.edu/abs/2021orel.bookE...1V}
}

@ARTICLE{Ronco2024,
       author = {{Ronco}, M.~P. and {Schreiber}, M.~R. and {Villaver}, E. and {Guilera}, O.~M. and {Miller Bertolami}, M.~M.},
        title = "{Planet formation around intermediate-mass stars. I. Different disc evolutionary pathways as a function of stellar mass}",
      journal = {\aap},
         year = 2024,
        month = feb,
       volume = {682},
          eid = {A155},
        pages = {A155},
          doi = {10.1051/0004-6361/202347762},
archivePrefix = {arXiv},
       eprint = {2311.03934},
 primaryClass = {astro-ph.EP},
       adsurl = {https://ui.adsabs.harvard.edu/abs/2024A&A...682A.155R}
}

@ARTICLE{Lodders2025,
       author = {{Lodders}, K. and {Bergemann}, M. and {Palme}, H.},
        title = "{Solar System Elemental Abundances from the Solar Photosphere and CI-Chondrites}",
      journal = {\ssr},
         year = 2025,
        month = mar,
       volume = {221},
       number = {2},
          eid = {23},
        pages = {23},
          doi = {10.1007/s11214-025-01146-w},
archivePrefix = {arXiv},
       eprint = {2502.10575},
 primaryClass = {astro-ph.SR},
       adsurl = {https://ui.adsabs.harvard.edu/abs/2025SSRv..221...23L}
}

@article{Lambrechts2014,
	Adsurl = {http://adsabs.harvard.edu/abs/2014A%26A...572A..35L},
	Archiveprefix = {arXiv},
	Author = {{Lambrechts}, M. and {Johansen}, A. and {Morbidelli}, A.},
	Doi = {10.1051/0004-6361/201423814},
	Eid = {A35},
	Eprint = {1408.6087},
	Journal = {\aap},
	Month = dec,
	Pages = {A35},
	Primaryclass = {astro-ph.EP},
	Title = {{Separating gas-giant and ice-giant planets by halting pebble accretion}},
	Volume = 572,
	Year = 2014}

@article{LambrechtsJohansen2014,
	Adsurl = {http://adsabs.harvard.edu/abs/2014A%26A...572A.107L},
	Archiveprefix = {arXiv},
	Author = {{Lambrechts}, M. and {Johansen}, A.},
	Doi = {10.1051/0004-6361/201424343},
	Eid = {A107},
	Eprint = {1408.6094},
	Journal = {\aap},
	Month = dec,
	Pages = {A107},
	Primaryclass = {astro-ph.EP},
	Title = {{Forming the cores of giant planets from the radial pebble flux in protoplanetary discs}},
	Volume = 572,
	Year = 2014}

@ARTICLE{Venturini2024,
       author = {{Venturini}, J. and {Ronco}, M.~P. and {Guilera}, O.~M. and {Haldemann}, J. and {Mordasini}, C. and {Miller Bertolami}, M.},
        title = "{A fading radius valley towards M dwarfs, a persistent density valley across stellar types}",
      journal = {\aap},
         year = 2024,
        month = jun,
       volume = {686},
          eid = {L9},
        pages = {L9},
          doi = {10.1051/0004-6361/202349088},
archivePrefix = {arXiv},
       eprint = {2404.01967},
 primaryClass = {astro-ph.EP},
       adsurl = {https://ui.adsabs.harvard.edu/abs/2024A&A...686L...9V}
}

@ARTICLE{LiuBeibei2019,
       author = {{Liu}, Beibei and {Lambrechts}, Michiel and {Johansen}, Anders and {Liu}, Fan},
        title = "{Super-Earth masses sculpted by pebble isolation around stars of different masses}",
      journal = {\aap},
         year = 2019,
        month = dec,
       volume = {632},
          eid = {A7},
        pages = {A7},
          doi = {10.1051/0004-6361/201936309},
archivePrefix = {arXiv},
       eprint = {1909.00759},
 primaryClass = {astro-ph.EP},
       adsurl = {https://ui.adsabs.harvard.edu/abs/2019A&A...632A...7L}
}

@ARTICLE{Kanagawa2018,
       author = {{Kanagawa}, Kazuhiro D. and {Tanaka}, Hidekazu and {Szuszkiewicz}, Ewa},
        title = "{Radial Migration of Gap-opening Planets in Protoplanetary Disks. I. The Case of a Single Planet}",
      journal = {\apj},
         year = 2018,
        month = jul,
       volume = {861},
       number = {2},
          eid = {140},
        pages = {140},
          doi = {10.3847/1538-4357/aac8d9},
archivePrefix = {arXiv},
       eprint = {1805.11101},
 primaryClass = {astro-ph.EP},
       adsurl = {https://ui.adsabs.harvard.edu/abs/2018ApJ...861..140K}
}

@ARTICLE{Ida2018,
       author = {{Ida}, Shigeru and {Tanaka}, Hidekazu and {Johansen}, Anders and {Kanagawa}, Kazuhiro D. and {Tanigawa}, Takayuki},
        title = "{Slowing Down Type II Migration of Gas Giants to Match Observational Data}",
      journal = {\apj},
         year = 2018,
        month = sep,
       volume = {864},
       number = {1},
          eid = {77},
        pages = {77},
          doi = {10.3847/1538-4357/aad69c},
archivePrefix = {arXiv},
       eprint = {1807.10871},
 primaryClass = {astro-ph.EP},
       adsurl = {https://ui.adsabs.harvard.edu/abs/2018ApJ...864...77I}
}

@ARTICLE{Wittenmyer2020,
       author = {{Wittenmyer}, Robert A. and {Butler}, R.~P. and {Horner}, Jonathan and {Clark}, Jake and {Tinney}, C.~G. and {Carter}, B.~D. and {Wang}, Liang and {Johnson}, John Asher and {Collins}, Michaela},
        title = "{The Pan-Pacific Planet Search - VIII. Complete results and the occurrence rate of planets around low-luminosity giants}",
      journal = {\mnras},
         year = 2020,
        month = feb,
       volume = {491},
       number = {4},
        pages = {5248-5257},
          doi = {10.1093/mnras/stz3378},
archivePrefix = {arXiv},
       eprint = {1911.11954},
 primaryClass = {astro-ph.EP},
       adsurl = {https://ui.adsabs.harvard.edu/abs/2020MNRAS.491.5248W}
}

@ARTICLE{Preibisch2005,
       author = {{Preibisch}, Thomas and {Feigelson}, Eric D.},
        title = "{The Evolution of X-Ray Emission in Young Stars}",
      journal = {\apjs},
         year = 2005,
        month = oct,
       volume = {160},
       number = {2},
        pages = {390-400},
          doi = {10.1086/432094},
archivePrefix = {arXiv},
       eprint = {astro-ph/0506052},
 primaryClass = {astro-ph},
       adsurl = {https://ui.adsabs.harvard.edu/abs/2005ApJS..160..390P}
}

@ARTICLE{VelascoRomero2020,
       author = {{Velasco Romero}, David A. and {Masset}, Fr{\'e}d{\'e}ric S.},
        title = "{Dynamical friction with radiative feedback - II. High-resolution study of the subsonic regime}",
      journal = {\mnras},
         year = 2020,
        month = jun,
       volume = {495},
       number = {2},
        pages = {2063-2074},
          doi = {10.1093/mnras/staa1215},
archivePrefix = {arXiv},
       eprint = {2004.13422},
 primaryClass = {astro-ph.EP},
       adsurl = {https://ui.adsabs.harvard.edu/abs/2020MNRAS.495.2063V}
}

@ARTICLE{WinterHaworth2022,
       author = {{Winter}, Andrew J. and {Haworth}, Thomas J.},
        title = "{The external photoevaporation of planet-forming discs}",
      journal = {European Physical Journal Plus},
         year = 2022,
        month = oct,
       volume = {137},
       number = {10},
          eid = {1132},
        pages = {1132},
          doi = {10.1140/epjp/s13360-022-03314-1},
archivePrefix = {arXiv},
       eprint = {2206.11910},
 primaryClass = {astro-ph.EP},
       adsurl = {https://ui.adsabs.harvard.edu/abs/2022EPJP..137.1132W}
}

@ARTICLE{Mauch-Soriano2026,
       author = {{Mauch-Soriano}, Alex and {Schreiber}, Matthias R. and {Correa}, Diego and {Pinilla}, Julio and {Riveros-Jara}, Catalina and {Vivanco}, Javiera and {Ronco}, Maria Paula and {Belloni}, Diogo and {Lagos-Vilches}, Felipe and {Brandner}, Wolfgang},
        title = "{Predicted incidence of Jupiter-like planets around white dwarfs}",
      journal = {\aap},
         year = 2026,
        month = mar,
       volume = {707},
          eid = {A90},
        pages = {A90},
          doi = {10.1051/0004-6361/202558367},
archivePrefix = {arXiv},
       eprint = {2601.19531},
 primaryClass = {astro-ph.EP},
       adsurl = {https://ui.adsabs.harvard.edu/abs/2026A&A...707A..90M}
}

@ARTICLE{Marigo2020,
       author = {{Marigo}, Paola and {Cummings}, Jeffrey D. and {Curtis}, Jason Lee and {Kalirai}, Jason and {Chen}, Yang and {Tremblay}, Pier-Emmanuel and {Ramirez-Ruiz}, Enrico and {Bergeron}, Pierre and {Bladh}, Sara and {Bressan}, Alessandro and {Girardi}, L{\'e}o and {Pastorelli}, Giada and {Trabucchi}, Michele and {Cheng}, Sihao and {Aringer}, Bernhard and {Tio}, Piero Dal},
        title = "{Carbon star formation as seen through the non-monotonic initial-final mass relation}",
      journal = {Nature Astronomy},
         year = 2020,
        month = jul,
       volume = {4},
        pages = {1102-1110},
          doi = {10.1038/s41550-020-1132-1},
archivePrefix = {arXiv},
       eprint = {2007.04163},
 primaryClass = {astro-ph.SR},
       adsurl = {https://ui.adsabs.harvard.edu/abs/2020NatAs...4.1102M}
}

@ARTICLE{Cornejo2023,
       author = {{Cornejo}, Sonia and {Masset}, Fr{\'e}d{\'e}ric S. and {S{\'a}nchez-Salcedo}, F.~J.},
        title = "{On the interaction of pebble accreting embryos with the gaseous disc: importance of thermal forces}",
      journal = {\mnras},
         year = 2023,
        month = jul,
       volume = {523},
       number = {1},
        pages = {936-953},
          doi = {10.1093/mnras/stad1476},
archivePrefix = {arXiv},
       eprint = {2305.09093},
 primaryClass = {astro-ph.EP},
       adsurl = {https://ui.adsabs.harvard.edu/abs/2023MNRAS.523..936C}
}

@ARTICLE{Masset2017MNRAS,
       author = {{Masset}, Fr{\'e}d{\'e}ric S.},
        title = "{Coorbital thermal torques on low-mass protoplanets}",
      journal = {\mnras},
         year = 2017,
        month = dec,
       volume = {472},
       number = {4},
        pages = {4204-4219},
          doi = {10.1093/mnras/stx2271},
archivePrefix = {arXiv},
       eprint = {1708.09807},
 primaryClass = {astro-ph.EP},
       adsurl = {https://ui.adsabs.harvard.edu/abs/2017MNRAS.472.4204M}
}

@ARTICLE{Eklund2017MNRAS,
       author = {{Eklund}, Henrik and {Masset}, Fr{\'e}d{\'e}ric S.},
        title = "{Evolution of eccentricity and inclination of hot protoplanets embedded in radiative discs}",
      journal = {\mnras},
         year = 2017,
        month = jul,
       volume = {469},
       number = {1},
        pages = {206-217},
          doi = {10.1093/mnras/stx856},
archivePrefix = {arXiv},
       eprint = {1704.01931},
 primaryClass = {astro-ph.EP},
       adsurl = {https://ui.adsabs.harvard.edu/abs/2017MNRAS.469..206E}
}

@ARTICLE{Fromenteau2019MNRAS,
       author = {{Fromenteau}, S{\'e}bastien and {Masset}, Fr{\'e}d{\'e}ric S.},
        title = "{Impact of thermal effects on the evolution of eccentricity and inclination of low-mass planets}",
      journal = {\mnras},
         year = 2019,
        month = jun,
       volume = {485},
       number = {4},
        pages = {5035-5049},
          doi = {10.1093/mnras/stz718},
archivePrefix = {arXiv},
       eprint = {1903.04470},
 primaryClass = {astro-ph.EP},
       adsurl = {https://ui.adsabs.harvard.edu/abs/2019MNRAS.485.5035F}
}

@ARTICLE{Jones2015,
       author = {{Jones}, M.~I. and {Jenkins}, J.~S. and {Rojo}, P. and {Olivares}, F. and {Melo}, C.~H.~F.},
        title = "{Giant planets around two intermediate-mass evolved stars and confirmation of the planetary nature of HIP 67851c}",
      journal = {\aap},
         year = 2015,
        month = aug,
       volume = {580},
          eid = {A14},
        pages = {A14},
          doi = {10.1051/0004-6361/201525853},
archivePrefix = {arXiv},
       eprint = {1505.06718},
 primaryClass = {astro-ph.EP},
       adsurl = {https://ui.adsabs.harvard.edu/abs/2015A&A...580A..14J}
}

@ARTICLE{Mullally2024,
       author = {{Mullally}, Susan E. and {Debes}, John and {Cracraft}, Misty and {Mullally}, Fergal and {Poulsen}, Sabrina and {Albert}, Loic and {Thibault}, Katherine and {Reach}, William T. and {Hermes}, J.~J. and {Barclay}, Thomas and {Kilic}, Mukremin and {Quintana}, Elisa V.},
        title = "{JWST Directly Images Giant Planet Candidates Around Two Metal-polluted White Dwarf Stars}",
      journal = {\apjl},
         year = 2024,
        month = feb,
       volume = {962},
       number = {2},
          eid = {L32},
        pages = {L32},
          doi = {10.3847/2041-8213/ad2348},
archivePrefix = {arXiv},
       eprint = {2401.13153},
 primaryClass = {astro-ph.EP},
       adsurl = {https://ui.adsabs.harvard.edu/abs/2024ApJ...962L..32M}
}

@ARTICLE{Limbach2024,
       author = {{Limbach}, Mary Anne and {Vanderburg}, Andrew and {Venner}, Alexander and {Blouin}, Simon and {Stevenson}, Kevin B. and {MacDonald}, Ryan J. and {Jenkins}, Sydney and {Bowens-Rubin}, Rachel and {Soares-Furtado}, Melinda and {Morley}, Caroline and {Janson}, Markus and {Debes}, John and {Xu}, Siyi and {Kleisioti}, Evangelia and {Kenworthy}, Matthew and {Butler}, Paul and {Crane}, Jeffrey D. and {Osip}, Dave and {Shectman}, Stephen and {Teske}, Johanna},
        title = "{The MIRI Exoplanets Orbiting White dwarfs (MEOW) Survey: Mid-infrared Excess Reveals a Giant Planet Candidate around a Nearby White Dwarf}",
      journal = {\apjl},
         year = 2024,
        month = sep,
       volume = {973},
       number = {1},
          eid = {L11},
        pages = {L11},
          doi = {10.3847/2041-8213/ad74ed},
archivePrefix = {arXiv},
       eprint = {2408.16813},
 primaryClass = {astro-ph.EP},
       adsurl = {https://ui.adsabs.harvard.edu/abs/2024ApJ...973L..11L}
}

@ARTICLE{Stapper2026,
       author = {{Stapper}, L.~M. and {Vioque}, M. and {Winter}, A.~J. and {Wilner}, D.~J. and {Williams}, J.~P. and {Benisty}, M. and {Booth}, A.~S. and {Grant}, S.~L. and {Hogerheijde}, M.~R.},
        title = "{An analysis of the Herbig star population and their protoplanetary disks within 1 kpc}",
      journal = {arXiv e-prints},
         year = 2026,
        month = aug,
          eid = {arXiv:2608.23729},
        pages = {arXiv:2608.23729},
          doi = {10.48550/arXiv.2608.23729},
archivePrefix = {arXiv},
       eprint = {2608.23729},
 primaryClass = {astro-ph.EP},
       adsurl = {https://ui.adsabs.harvard.edu/abs/2026arXiv260823729S}
}

@ARTICLE{Pan2025,
       author = {{Pan}, Mengrui and {Liu}, Beibei and {Jiang}, Linjie and {Xie}, Jiwei and {Zhu}, Wei and {Ribas}, Ignasi},
        title = "{Dependence of Planet Populations on Stellar Mass and Metallicity: A Pebble-accretion-based Planet Population Synthesis Model}",
      journal = {\apj},
         year = 2025,
        month = may,
       volume = {985},
       number = {1},
          eid = {7},
        pages = {7},
          doi = {10.3847/1538-4357/adc7a9},
archivePrefix = {arXiv},
       eprint = {2504.00296},
 primaryClass = {astro-ph.EP},
       adsurl = {https://ui.adsabs.harvard.edu/abs/2025ApJ...985....7P}
}

@ARTICLE{Mulders2021,
       author = {{Mulders}, Gijs D. and {Dr{\k{a}}{\.z}kowska}, Joanna and {van der Marel}, Nienke and {Ciesla}, Fred J. and {Pascucci}, Ilaria},
        title = "{Why Do M Dwarfs Have More Transiting Planets?}",
      journal = {\apjl},
         year = 2021,
        month = oct,
       volume = {920},
       number = {1},
          eid = {L1},
        pages = {L1},
          doi = {10.3847/2041-8213/ac2947},
archivePrefix = {arXiv},
       eprint = {2110.02971},
 primaryClass = {astro-ph.EP},
       adsurl = {https://ui.adsabs.harvard.edu/abs/2021ApJ...920L...1M}
}

@ARTICLE{Maldonado2025,
       author = {{Maldonado}, J. and {Mirouh}, G.~M. and {Mendigut{\'\i}a}, I. and {Montesinos}, B. and {Gragera-M{\'a}s}, J.~L. and {Villaver}, E.},
        title = "{Intermediate-mass stars and the origin of the gas-giant planet-metallicity correlation}",
      journal = {\aap},
         year = 2025,
        month = mar,
       volume = {695},
          eid = {A27},
        pages = {A27},
          doi = {10.1051/0004-6361/202453328},
archivePrefix = {arXiv},
       eprint = {2501.19074},
 primaryClass = {astro-ph.SR},
       adsurl = {https://ui.adsabs.harvard.edu/abs/2025A&A...695A..27M}
}

@ARTICLE{Maldonado2026,
       author = {{Maldonado}, J. and {Mendigut{\'\i}a}, I. and {Barcel{\'o} Forteza}, S. and {Mirouh}, G.~M. and {Montesinos}, B. and {Villaver}, E.},
        title = "{Signatures of rocky debris accretion in AF-type planet hosts}",
      journal = {\aap},
         year = 2026,
        month = oct,
       volume = {714},
          eid = {A265},
        pages = {A265},
          doi = {10.1051/0004-6361/202661077},
archivePrefix = {arXiv},
       eprint = {2609.04790},
 primaryClass = {astro-ph.EP},
       adsurl = {https://ui.adsabs.harvard.edu/abs/2026A&A...714A.265M}
}

@ARTICLE{Jones2011,
       author = {{Jones}, M.~I. and {Jenkins}, J.~S. and {Rojo}, P. and {Melo}, C.~H.~F.},
        title = "{Study of the impact of the post-MS evolution of the host star on the orbits of close-in planets. I. Sample definition and physical properties}",
      journal = {\aap},
         year = 2011,
        month = dec,
       volume = {536},
          eid = {A71},
        pages = {A71},
          doi = {10.1051/0004-6361/201117887},
archivePrefix = {arXiv},
       eprint = {1110.6459},
 primaryClass = {astro-ph.EP},
       adsurl = {https://ui.adsabs.harvard.edu/abs/2011A&A...536A..71J}
}

@ARTICLE{Wittenmyer2011,
       author = {{Wittenmyer}, Robert A. and {Endl}, Michael and {Wang}, Liang and {Johnson}, John Asher and {Tinney}, C.~G. and {O'Toole}, S.~J.},
        title = "{The Pan-Pacific Planet Search. I. A Giant Planet Orbiting 7 CMa}",
      journal = {\apj},
         year = 2011,
        month = dec,
       volume = {743},
       number = {2},
          eid = {184},
        pages = {184},
          doi = {10.1088/0004-637X/743/2/184},
archivePrefix = {arXiv},
       eprint = {1111.1007},
 primaryClass = {astro-ph.EP},
       adsurl = {https://ui.adsabs.harvard.edu/abs/2011ApJ...743..184W}
}

@ARTICLE{Chomez2025,
       author = {{Chomez}, A. and {Delorme}, P. and {Lagrange}, A.-M. and {Gratton}, R. and {Flasseur}, O. and {Chauvin}, G. and {Langlois}, M. and {Mazoyer}, J. and {Zurlo}, A. and {Desidera}, S. and {Mesa}, D. and {Bonnefoy}, M. and {Feldt}, M. and {Hagelberg}, J. and {Meyer}, M. and {Vigan}, A. and {Ginski}, C. and {Kenworthy}, M. and {Albert}, D. and {Bergeon}, S. and {Beuzit}, J.-L. and {Biller}, B. and {Bhowmik}, T. and {Boccaletti}, A. and {Bonavita}, M. and {Brandner}, W. and {Cantalloube}, F. and {Cheetham}, A. and {D'Orazi}, V. and {Dominik}, C. and {Fontanive}, C. and {Galicher}, R. and {Henning}, Th. and {Janson}, M. and {Kral}, Q. and {Lagadec}, E. and {Lazzoni}, C. and {Le Coroller}, H. and {Ligi}, R. and {Maire}, A.-L. and {Marleau}, G.-D. and {Menard}, F. and {Messina}, S. and {Meunier}, N. and {Mordasini}, C. and {Moutou}, C. and {M{\"u}ller}, A. and {Perrot}, C. and {Samland}, M. and {Schmid}, H.~M. and {Schmidt}, T. and {Squicciarini}, V. and {Sissa}, E. and {Turatto}, M. and {Udry}, S. and {Abe}, L. and {Antichi}, J. and {Asensio-Torres}, R. and {Baruffolo}, A. and {Baudoz}, P. and {Baudrand}, J. and {Bazzon}, A. and {Blanchard}, P. and {Bohn}, A.~J. and {Brown Sevilla}, S. and {Carbillet}, M. and {Carle}, M. and {Cascone}, E. and {Charton}, J. and {Claudi}, R. and {Costille}, A. and {De Caprio}, V. and {Delboulb{\'e}}, A. and {Dohlen}, K. and {Engler}, N. and {Fantinel}, D. and {Feautrier}, P. and {Fusco}, T. and {Gigan}, P. and {Girard}, J.~H. and {Giro}, E. and {Gisler}, D. and {Gluck}, L. and {Gry}, C. and {Hubin}, N. and {Hugot}, E. and {Jaquet}, M. and {Kasper}, M. and {Le Mignant}, D. and {Llored}, M. and {Madec}, F. and {Magnard}, Y. and {Martinez}, P. and {Maurel}, D. and {M{\"o}ller-Nilsson}, O. and {Mouillet}, D. and {Moulin}, T. and {Orign{\'e}}, A. and {Pavlov}, A. and {Perret}, D. and {Petit}, C. and {Pragt}, J. and {Puget}, P. and {Rabou}, P. and {Ramos}, J. and {Rickman}, E.~L. and {Rigal}, F. and {Rochat}, S. and {Roelfsema}, R. and {Rousset}, G. and {Roux}, A. and {Salasnich}, B. and {Sauvage}, J.-F. and {Sevin}, A. and {Soenke}, C. and {Stadler}, E. and {Suarez}, M. and {Wahhaj}, Z. and {Weber}, L. and {Wildi}, F.},
        title = "{The SPHERE infrared survey for exoplanets (SHINE): IV. Complete observations, data reduction and analysis, detection performances, and final results}",
      journal = {\aap},
         year = 2025,
        month = may,
       volume = {697},
          eid = {A99},
        pages = {A99},
          doi = {10.1051/0004-6361/202451751},
archivePrefix = {arXiv},
       eprint = {2501.12002},
 primaryClass = {astro-ph.EP},
       adsurl = {https://ui.adsabs.harvard.edu/abs/2025A&A...697A..99C}
}

@ARTICLE{Delorme2024,
       author = {{Delorme}, P. and {Chomez}, A. and {Squicciarini}, V. and {Janson}, M. and {Flasseur}, O. and {Schib}, O. and {Gratton}, R. and {Lagrange}, A.-M. and {Langlois}, M. and {Mayer}, L. and {Helled}, R. and {Reffert}, S. and {Kiefer}, F. and {Biller}, B. and {Chauvin}, G. and {Fontanive}, C. and {Henning}, Th. and {Kenworthy}, M. and {Marleau}, G.-D. and {Mesa}, D. and {Meyer}, M.~R. and {Mordasini}, C. and {Ringqvist}, S.~C. and {Samland}, M. and {Vigan}, A. and {Viswanath}, G.},
        title = "{Population of giant planets around B stars from the first part of the BEAST survey}",
      journal = {\aap},
         year = 2024,
        month = dec,
       volume = {692},
          eid = {A263},
        pages = {A263},
          doi = {10.1051/0004-6361/202451461},
archivePrefix = {arXiv},
       eprint = {2409.18793},
 primaryClass = {astro-ph.EP},
       adsurl = {https://ui.adsabs.harvard.edu/abs/2024A&A...692A.263D}
}

@ARTICLE{Jones2016,
       author = {{Jones}, M.~I. and {Jenkins}, J.~S. and {Brahm}, R. and {Wittenmyer}, R.~A. and {Olivares E.}, F. and {Melo}, C.~H.~F. and {Rojo}, P. and {Jord{\'a}n}, A. and {Drass}, H. and {Butler}, R.~P. and {Wang}, L.},
        title = "{Four new planets around giant stars and the mass-metallicity correlation of planet-hosting stars}",
      journal = {\aap},
         year = 2016,
        month = may,
       volume = {590},
          eid = {A38},
        pages = {A38},
          doi = {10.1051/0004-6361/201628067},
archivePrefix = {arXiv},
       eprint = {1603.03738},
 primaryClass = {astro-ph.EP},
       adsurl = {https://ui.adsabs.harvard.edu/abs/2016A&A...590A..38J}
}

@ARTICLE{Frink2001,
       author = {{Frink}, Sabine and {Quirrenbach}, Andreas and {Fischer}, Debra and {R{\"o}ser}, Siegfried and {Schilbach}, Elena},
        title = "{A Strategy for Identifying the Grid Stars for the Space Interferometry Mission}",
      journal = {\pasp},
         year = 2001,
        month = feb,
       volume = {113},
       number = {780},
        pages = {173-187},
          doi = {10.1086/318610},
archivePrefix = {arXiv},
       eprint = {astro-ph/0011190},
 primaryClass = {astro-ph},
       adsurl = {https://ui.adsabs.harvard.edu/abs/2001PASP..113..173F}
}

@ARTICLE{Miranda2026,
       author = {{Miranda}, Nicolas F. and {Mauch-Soriano}, Alex and {Barraza-Jorquera}, Joaquin A. and {Schreiber}, Matthias R. and {Ronco}, Maria Paula and {Lagos-Vilches}, Felipe},
        title = "{Population synthesis of substellar companions surviving common-envelope evolution}",
      journal = {arXiv e-prints},
         year = 2026,
        month = nov,
          eid = {arXiv:2611.06549},
        pages = {arXiv:2611.06549},
          doi = {10.48550/arXiv.2611.06549},
archivePrefix = {arXiv},
       eprint = {2611.06549},
 primaryClass = {astro-ph.SR},
       adsurl = {https://ui.adsabs.harvard.edu/abs/2026arXiv261106549M}
}

\begin{appendix}

\section{The effects of thermal torques}
\label{Ap:TT}

\begin{figure*}[ht!]
  \centering
    \includegraphics[angle=0, width= \textwidth]{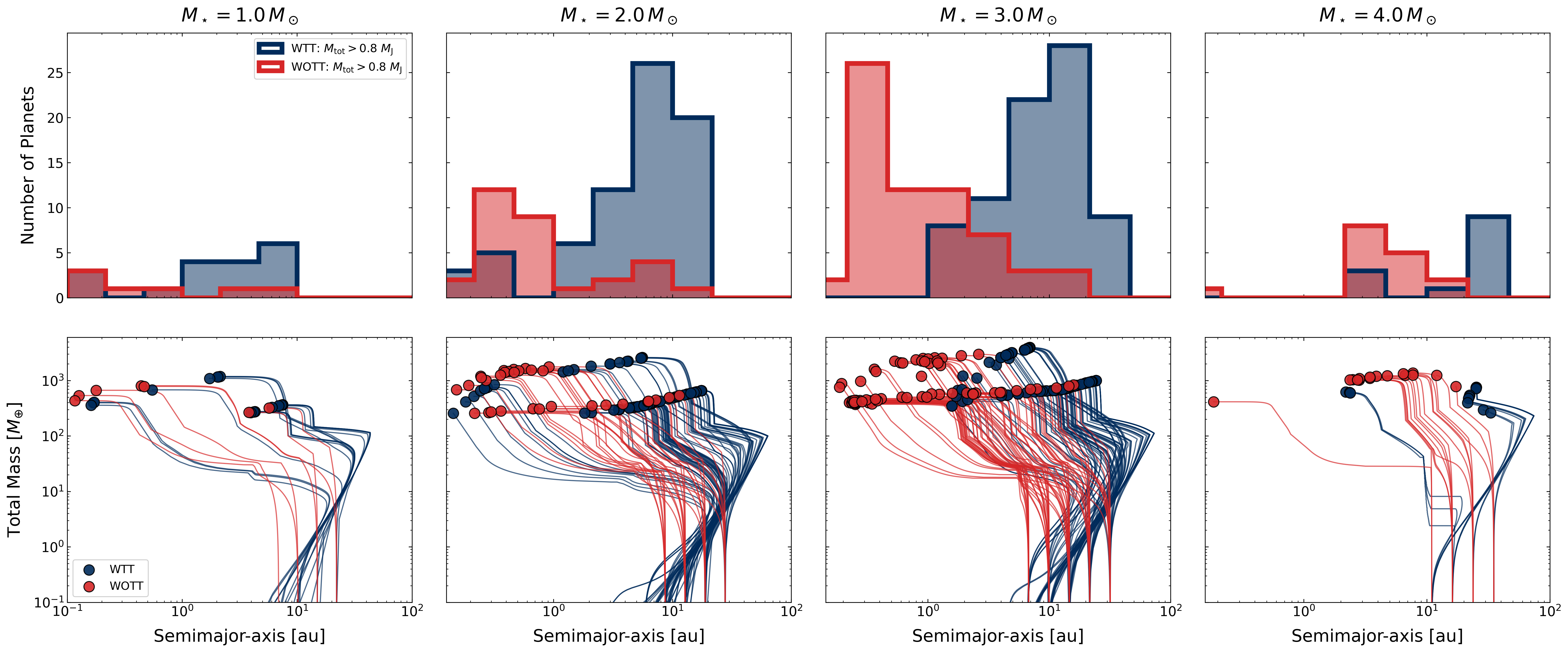}
    \caption{Semimajor axis distributions (top panel) and planet formation tracks for planets reaching final masses greater than $0.8M_\text{J}$ in the high disk-mass scenarios for $1\mathrm{M}_\odot$, $2\mathrm{M}_\odot$, $3\mathrm{M}_\odot$, and $4\mathrm{M}_\odot$ stars. Red and blue histograms and tracks represent simulations without thermal torques (WOTT) and with thermal torques (WTT), respectively.}
\label{Fig:Ap-Thermal-Torques}
\end{figure*}

Figure \ref{Fig:Ap-Thermal-Torques} shows the semimajor axis distribution (top panel) and the planet formation tracks exclusively for planets with final masses greater than $0.8M_\text{J}$, focusing on the high disk-mass scenarios for $1\mathrm{M}_\odot$, $2\mathrm{M}_\odot$, $3\mathrm{M}_\odot$, and $4\mathrm{M}_\odot$. The red and blue histograms, along with their corresponding formation tracks and final locations, represent simulations without thermal torques (WOTT) and with thermal torques (WTT), respectively. Two primary differences emerge between these setups: the WTT simulations produce a higher number of gas giant planets compared to the WOTT cases, and the final orbital locations of these giants differ significantly. In the $1\mathrm{M}_\odot$, $2\mathrm{M}_\odot$, and $3\mathrm{M}_\odot$ scenarios, the WTT models produce 2.57, 2.32, and 1.2 times as many giant planets as the WOTT models, respectively. However, this trend reverses in the $4\mathrm{M}_\odot$ case, where the WTT setup yields only 0.81 times as many giants.

As previously mentioned, giants in the WTT models tend to accumulate beyond a few au, whereas those in the WOTT models typically migrate toward the inner regions in the disk. An exception occurs in the $4\mathrm{M}_\odot$ case: driven by rapid outside-in disk evolution, planets in both scenarios remain beyond 2 au (although they orbit even farther out in the WTT case). Furthermore, both models preserve the trend of forming more massive planets as stellar mass increases, clearly showing that this correlation is not an effect of thermal torques.

It is important to highlight that the adopted recipes from \citet{Masset2017MNRAS} and \citet{VelascoRomero2020} to compute the thermal torques do not account for the growth of planetary eccentricity driven by thermal forces, particularly the heating component. Previous studies by \citet{Eklund2017MNRAS} and \citet{Fromenteau2019MNRAS} have shown that the heating torque can indeed excite the eccentricity of a planet. Furthermore, \citet{Cornejo2023} demonstrated that this resulting increase in eccentricity can limit outward migration by damping the overall thermal torque. Consequently, a more comprehensive treatment of planet migration would likely yield formation tracks that fall somewhere between our WTT and WOTT scenarios.

\section{Disk evolution and planet formation with and without stellar evotution}
\label{App:Msyst}

\begin{figure*}[ht!]
  \centering
    \includegraphics[angle=0, width= \textwidth]{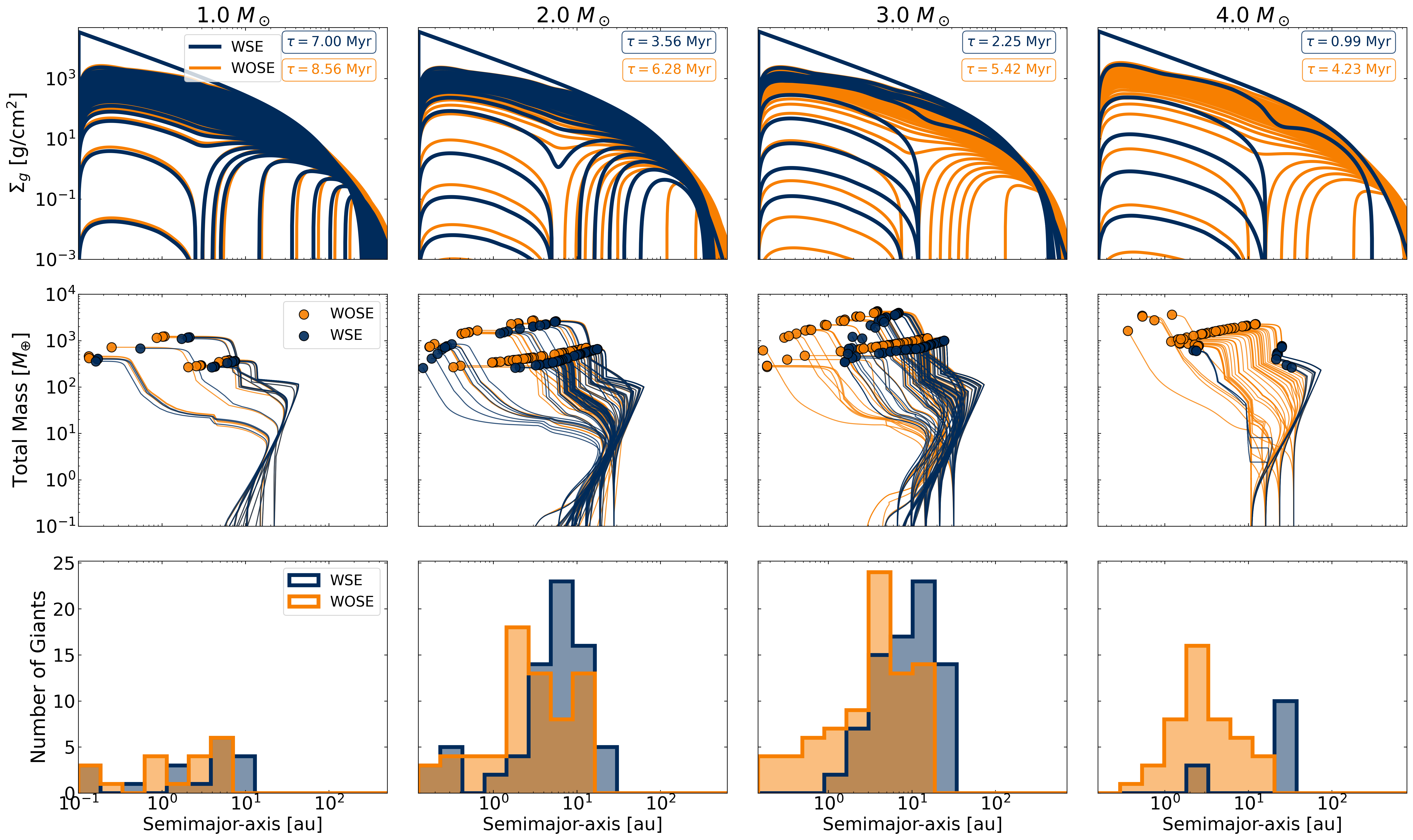}
    \caption{Comparison of gas disk evolution (top row), giant planet formation tracks (middle row), and final semimajor-axis distributions (bottom row) between models with (WSE) and without (WOSE) stellar evolution in the high disk-mass scenarios for $1\mathrm{M}_\odot$, $2\mathrm{M}_\odot$, $3\mathrm{M}_\odot$, and $4\mathrm{M}_\odot$ stars.}
\label{Fig:App-Comparacion}
\end{figure*}

The main goal of this work is to model planet formation around intermediate-mass stars, accounting for the effects of stellar evolution during the protoplanetary disk phase. In this section we show that this effect is crucial, as it alters not only the disk evolution pathways but also the planet formation outcomes. As previously described, we incorporate this effect primarily through time-dependent changes in the X-ray and FUV photoevaporation rates, driven by the temporal evolution of $L_X$ and $L_{\mathrm{FUV}}$ \citep{Kunitomo2021}. Previous models, such as that recently presented by \citet{johnston_2026}, incorporate evolving pre-main-sequence luminosities to compute the disk structure, yet they adopted a simplified approach regarding disk dispersal by assuming time-independent photoevaporation rates. Specifically, they implement X-ray photoevaporation rates scaled by stellar mass following the relation from \citet{Preibisch2005}. While this relation is primarily calibrated for stars with $M \lesssim 2 \, \mathrm{M}_\odot$, they extrapolate it across their entire studied mass range (up to $5\mathrm{M}_\odot$). Additionally, their framework focuses only on X-ray-driven mass loss, whereas our model expands upon this by including time-dependent FUV photoevaporation, which becomes a dominant mechanism for higher-mass stars, as shown by \citet{Kunitomo2021} and \citetalias{Ronco2024}.

Figure \ref{Fig:App-Comparacion} shows a comparison between our models with stellar evolution (WSE), focusing here only on the high-mass disk cases, and a set of simulations that assume $L_\text{X} \propto M_\star^{1.5}$ \citep{Preibisch2005}, as in \citet{johnston_2026}, and do not include FUV photoevaporation (WOSE). The top row compares the gas disk evolution between our case (WSE, in blue, and the same as in Figure \ref{Nominal-Case}) and the case without time dependent X-ray photoevaporation and FUV photoevaporation rates (WOSE, in orange). Although the disk dissipation timescale decreases significantly with increasing stellar mass in both scenarios, omitting the time-varying nature of $L_\text{X}$ and $L_\text{FUV}$ leads to substantial differences in the disk's evolutionary track, particularly for $M_\star \ge 3\mathrm{M}_\odot$. 
As a consequence, this strongly affects planet formation, as illustrated in the second row of Figure \ref{Fig:App-Comparacion}. Although the disk evolution and planetary outcomes for $1\,\mathrm{M}_\odot$ and $2\,\mathrm{M}_\odot$ do not differ significantly, the discrepancies in the $3\,\mathrm{M}_\odot$ and $4\,\mathrm{M}_\odot$ cases, are substantial. Both the faster disk dissipation and the shift in the dispersal mechanism, from an inside-out to an outside-in clearing, alter the final outcomes. In the WSE cases, the giant planets that form fail to migrate toward the inner regions of the disk. This happens because FUV photoevaporation drives a rapid gap opening, which quickly traps these planets in the outer regions, leaving them with insufficient time to migrate inward (most notably in the $4\,\mathrm{M}_\odot$ case). Conversely, in the WOSE scenario, planets have time to grow and migrate. Because the timescales for disk dissipation and gap opening are much longer, these processes occur only after the giant planet has already formed and reached its final mass and position. These differences are also clearly evident in the bottom row, which displays histograms of the number of planets as a function of their final position. It can be seen that for the $3\,\mathrm{M}_\odot$ and $4\,\mathrm{M}_\odot$ cases, the distributions change significantly. Giant planet formation is much more efficient in the WOSE case than in the WSE case as the mass of the star increases. Specifically, for the $1\,\mathrm{M}_\odot$ models, the WSE and WOSE cases yield 18 and 19 giant planets, respectively, showing a very similar semimajor axis distribution. A comparable trend is observed for $2\,\mathrm{M}_\odot$, with 72 and 69 giants formed, respectively. However, in the $3\,\mathrm{M}_\odot$ case, although the total number of giants remains similar (78 for WSE and 81 for WOSE), the orbital distribution begins to diverge, with a higher number of planets ending up closer to the star in the WOSE scenario. The most striking contrast occurs at $4\,\mathrm{M}_\odot$. Although the WOSE model produces 46 giant planets—distributed primarily between 1 and 10 au, the WSE model forms just 13 giants, all of which remain trapped beyond 10 au.

\section{Dependence on X-ray photoevaporation rates}
\label{Ap:PhotoRates}

\begin{figure*}[ht!]
 \centering
    \includegraphics[angle=0, width= \textwidth]{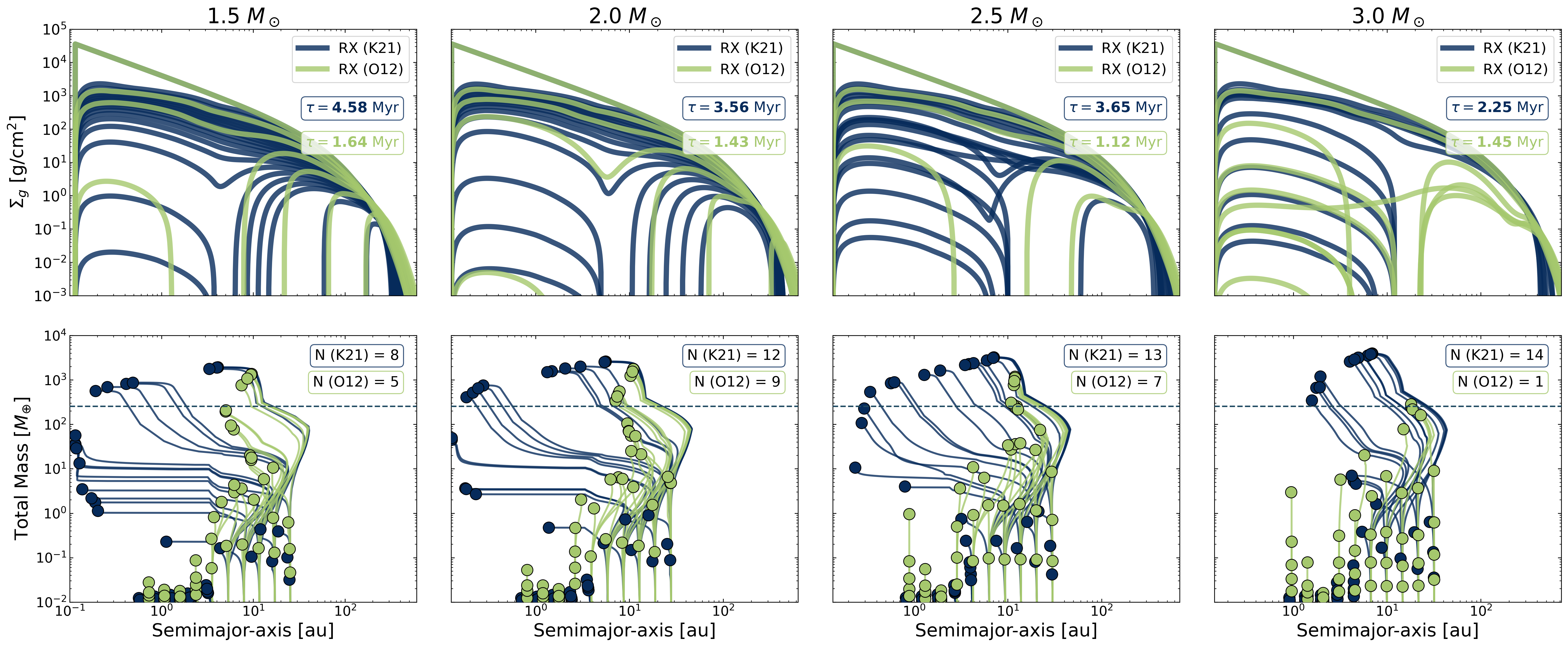}
    \caption{Comparison between disk evolution and planet formation adopting different X-ray photoevaporation rates. The top panel shows the time evolution of the gas surface density computed with the X-ray prescriptions for primordial disks by \citet{Kunitomo2021} (blue) and by \citet{Owen2012}. The bottom panel shows the planet formation tracks and final planet mass and locations for each corresponding case. The numbers represent the final number of gas giant planets formed in each case.}
\label{Fig:Ap-PhotoRates}
\end{figure*}

As discussed in Section \ref{Caveats}, the choice of X-ray photoevaporation prescription, combined with stellar evolution, significantly impacts both disk lifetimes and planetary outcomes. Figure \ref{Ap:PhotoRates} compares our nominal results, computed with the X-ray prescriptions of \citet{Kunitomo2021} for primordial disks (hereafter K21; see Fig. \ref{Nominal-Case}), with new simulations adopting the \citet{Owen2012} rates (hereafter O12). We focus on the $1.5-3 \mathrm{M}_\odot$ stellar mass range to capture both the observed peak in radial velocity  giant planet occurrence \citep{Wolthoff2022} and our theoretical peak, which lies near $3 \mathrm{M}_\odot$ (Section \ref{sec:impact_param}). The first row displays the gas surface density evolution for the K21 (blue) and O12 (green) models, while the second row shows the corresponding planet formation tracks and final outcomes. As explicitly labeled in each panel, disk dissipation timescales are substantially shorter in the O12 models than in the nominal K21 cases. This rapid dissipation severely restricts planetary growth, especially for gas giants, with the effect becoming more pronounced at higher stellar masses. Notably, the $3 \mathrm{M}_\odot$ disk in the O12 model takes more time to dissipate as it exhibits a revenant disk evolution (the same as Fig. 8 in \citetalias{Ronco2024}). Although the disk temporarily rebuilds after the initial X-ray-driven gap opens, slightly extending its total lifetime before FUV photoevaporation clears a second gap, this first gap opens very early (between $0.3$ and $0.4$ Myr). It is this initial, rapid gap opening that prematurely halts planetary growth. 

The bottom row of Figure \ref{Ap:PhotoRates} contrasts the planetary outcomes of the K21 models (blue) with those of the O12 models (green). The results diverge significantly: unlike in the K21 models, more planets in the outer disk of the O12 models fail to accrete enough mass to become gas giants. The few that do reach giant status become stranded in the outer disk; the rapid opening of the photoevaporation gap leaves these planets with insufficient time to migrate inward. This is particularly evident around $2.5 \mathrm{M}_\odot$ and $3 \mathrm{M}_\odot$ stars. As indicated in the panels, the K21 models show a steady increase in the number of gas giants with stellar mass (forming 8, 12, 13, and 14 giants around $1.5$, $2.0$, $2.5$, and $3.0 \mathrm{M}_\odot$, respectively). In contrast, the O12 models peak much earlier: the number of giants rises from 5 at $1.5 \mathrm{M}_\odot$ to 9 at $2.0 \mathrm{M}_\odot$, but then drops to 7 at $2.5 \mathrm{M}_\odot$ and to just a single gas giant at $3 \mathrm{M}_\odot$. Although we only simulate our nominal setup here, these findings suggest that more efficient X-ray photoevaporation rates, such as those of \citet{Owen2012}, naturally shift the theoretical peak of giant planet occurrence closer to $2 \mathrm{M}_\odot$. Consequently, they may be better suited than the \citet{Kunitomo2021} prescriptions for reproducing the RV occurrence rates reported by \citet{Wolthoff2022}.

\end{appendix}

\end{document}